\documentclass[sigplan,10pt]{acmart}

\renewcommand\footnotetextcopyrightpermission[1]{}

\usepackage[english]{babel}
\usepackage{blindtext}

\usepackage{tikz}
\usepackage{amsmath}
\usepackage{xspace}
\usepackage{url}
\usepackage{xurl}
\usepackage[most]{tcolorbox}
\usepackage{fancyhdr}
\usepackage{lineno}

\usepackage{listings}
\usepackage{xcolor}
\usepackage{caption}

\definecolor{codegreen}{rgb}{0,0.6,0}
\definecolor{codegray}{rgb}{0.5,0.5,0.5}
\definecolor{codepurple}{rgb}{0.58,0,0.82}
\definecolor{backcolour}{rgb}{0.98,0.98,0.98}

\fancypagestyle{firstpage}{
  \fancyhf{}
  \lhead{\includegraphics[height=20pt]{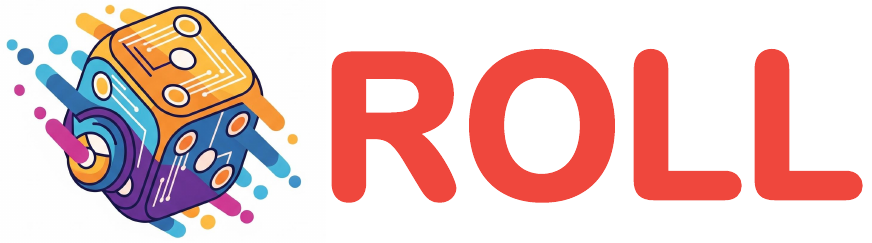}}
  \cfoot{\normalfont\footnotesize\thepage}

  \setlength{\headheight}{40pt}
}

\usepackage{balance}
\usepackage[normalem]{ulem}
\usepackage{graphicx}
\usepackage{algorithm}
\usepackage{algpseudocode}
\usepackage{amsthm}
\usepackage{amsmath}
\usepackage{multirow}
\usepackage{url}
\usepackage{subcaption}
\usepackage{enumitem}
\usepackage{makecell}
\usepackage{booktabs} 
\usepackage{soul}
\usepackage{textcomp}
\usepackage{pifont}
\usepackage{adjustbox}
\usepackage{wrapfig}

\newif\ifsubmit
\submitfalse 

\ifsubmit
  \newcommand{\appsecref}[1]{%
    \ifcsname appletter@#1\endcsname
      \csname appletter@#1\endcsname
    \else
      ??%
    \fi}
  \expandafter\def\csname appletter@app:rollout_profiling_thpt\endcsname{A}
  \expandafter\def\csname appletter@app:spot-instance-trace\endcsname{B}
  \expandafter\def\csname appletter@app:sensivity-to-lease\endcsname{C}
  \expandafter\def\csname appletter@app:sensivity-to-serving\endcsname{D}
  \expandafter\def\csname appletter@app:scalability\endcsname{E}
  \expandafter\def\csname appletter@app:timeline-breakdown\endcsname{F}
\else
  \newcommand{\appsecref}[1]{\ref*{#1}}
\fi

\usepackage{amssymb}
\usepackage{bbm}
\usepackage{tabularx}
\usepackage{bbding}

\usepackage{etoolbox}

\usepackage[normalem]{ulem}

\newcommand{\yuheng}[1]{\textcolor{black}{#1}}
\usepackage[most]{tcolorbox}

\usepackage{tikz}

\theoremstyle{definition}

\newcommand{\SystemName}{\textsc{ROSE}\xspace}

\microtypecontext{spacing=nonfrench}
\renewcommand\footnotetextcopyrightpermission[1]{} 
\acmDOI{}

\acmISBN{}
\begin{document}

\title[\SystemName: Rollout On Serving GPUs via Cooperative Elasticity for Agentic RL]
{\protect\vspace{0.5cm}\SystemName: \underline{R}ollout \underline{O}n \underline{S}erving GPUs via Cooperative \underline{E}lasticity for Agentic RL}



\author{
    \vspace{0.5cm}
    {\rm Wei Gao$^{\dagger*}$, Yuheng Zhao$^{\dagger*}$, Dilxat Muhtar$^{\ddagger}$, Dakai An$^{\dagger}$, Xuchun Shang$^{\ddagger}$,} \\
    {\rm Tianyuan Wu$^{\dagger}$, Lunxi Cao$^{\dagger}$, Shaopan Xiong$^{\ddagger}$, Weixun Wang$^{\ddagger}$, Ju Huang$^{\ddagger}$,} \\
    {\rm Teng Ma$^{\ddagger}$, Siran Yang$^{\ddagger}$, Jiamang Wang$^{\ddagger}$, Lin Qu$^{\ddagger}$, Bo Zheng$^{\ddagger}$, Wei Wang$^{\dagger}$}
    \bigskip \\
    {\em $^{\dagger}$HKUST} \hspace{2cm}
    {\em $^{\ddagger}$Alibaba Group} \\
    \smallskip
    {\small $^{*}$Equal contribution}
    \vspace{0.5cm}
}

\renewcommand{\shortauthors}{Gao, Zhao et al.}

\begin{abstract}

Agentic reinforcement learning (RL) is reshaping LLM post-training, but end-to-end training time is dominated by compute-intensive, multi-turn rollouts whose resource demand varies significantly across training steps. Resource-fixed systems cannot adapt to this variation, while resource-elastic approaches that provision external GPUs on demand suffer from high allocation overhead and limited availability.

We observe that serving clusters leave substantial GPU compute and memory idle, and propose \emph{cooperative elasticity}: sharing already-deployed serving GPUs with rollout workloads to provide on-demand elastic capacity. Realizing this is non-trivial, as it must preserve serving SLOs under bursty traffic while minimizing cross-cluster communication overhead. We present \SystemName{}, a system that realizes cooperative elasticity for agentic RL post-training, comprising three components: (1) an SLO-safe co-serving executor that co-locates heterogeneous serving and rollout models on the same GPUs, dynamically sharing memory and compute while preserving serving SLOs; (2) a cross-cluster weight transfer engine that leverages shard-aware routing and weight sparsity for fast synchronization; and (3) an elastic rollout scheduler that dynamically routes rollouts across dedicated and opportunistic serving GPUs. Experiments across multiple model sizes and cluster scales show that \SystemName{} improves end-to-end throughput by \(1.3\text{--}3.3\times\) over resource-fixed baselines and reduces rollout time by \(1.2\text{--}1.5\times\) over resource-elastic baselines, with no serving SLO violations.

\end{abstract}
\maketitle
\makeatletter
\g@addto@macro\@maketitle{\vspace{0.8cm}}
\makeatother
\thispagestyle{firstpage}

\pagestyle{plain}

\section{Introduction}

The advent of agentic reinforcement learning (RL) is reshaping large language model (LLM) post-training, shifting models from passive response generation to actively interacting with complex, dynamic environments. Recent advances in tool use~\cite{hao2025exploringsuperiorfunctioncalls,wu2025agenticreasoningstreamlinedframework}, computer use~\cite{luo2025guir1generalistr1style,lu2025uir1enhancingefficientaction,liu2025infigui}, and software engineering~\cite{swe-bench,tbench_2025,yang2025swe} suggest that this training paradigm enables LLMs to progressively solve complex tasks~\cite{li2025encouraginggoodprocessesneed,singh2025agenticreasoningtoolintegration,rawat2025preactmultistepplanningreasoning}. 
The standard agentic RL workflow alternates between two stages: \textit{rollout} and \textit{training}.
In the rollout stage, the agent LLM interacts with environments over multiple turns. At each turn, the agent \emph{decodes} tokens that serve as action signals, and then runs \emph{prefill} on the feedback returned by the environment. The resulting sequence of actions and feedback  forms a trajectory. In the training stage, the agent LLM updates its weights on the collected trajectories and synchronizes them to the rollout stage for the next iteration.  

In industry, LLM teams operate large training clusters for agentic RL and separate serving clusters for production deployment. Serving clusters are provisioned for peak traffic, but fluctuating demand often leaves GPUs underutilized~\cite{Lyra,lemix,BurstGPT,Prism,seallm,llm-multitasking,ConServe}. Meanwhile, agentic RL training imposes substantial resource demands, straining GPU capacity within training clusters.

\begin{figure*}[tb]
  \centering
    \begin{subfigure}{0.24\textwidth}
    \centering
    \includegraphics[width=\linewidth]{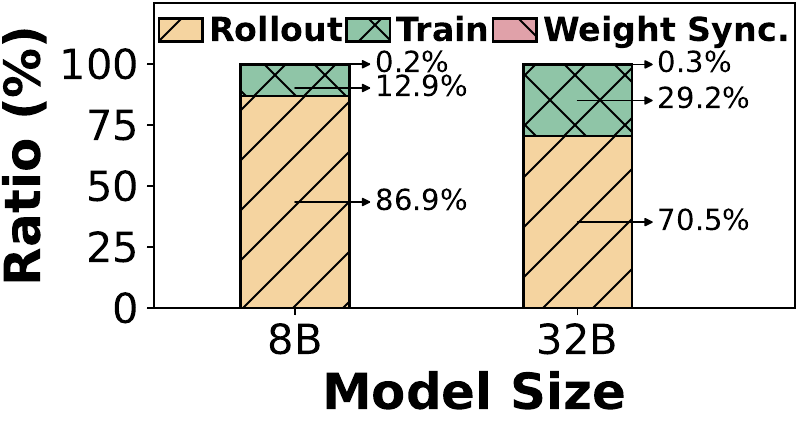}
    \caption{Performance Breakdown.}
    \label{subfig:perf-breakdown}
  \end{subfigure}\hfill
   \begin{subfigure}{0.24\textwidth}
    \centering
    \includegraphics[width=\linewidth]{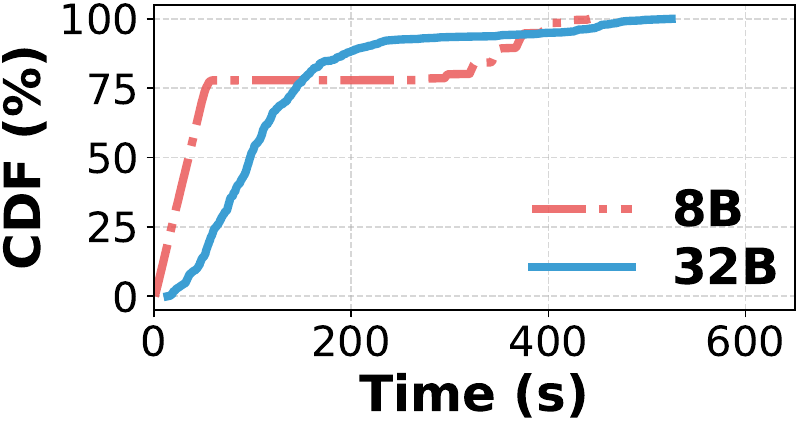}
    \caption{Long-tail Rollouts.}
    \label{subfig:long-tail}
  \end{subfigure}\hfill
  \begin{subfigure}{0.24\textwidth}
    \centering
    \includegraphics[width=\linewidth]{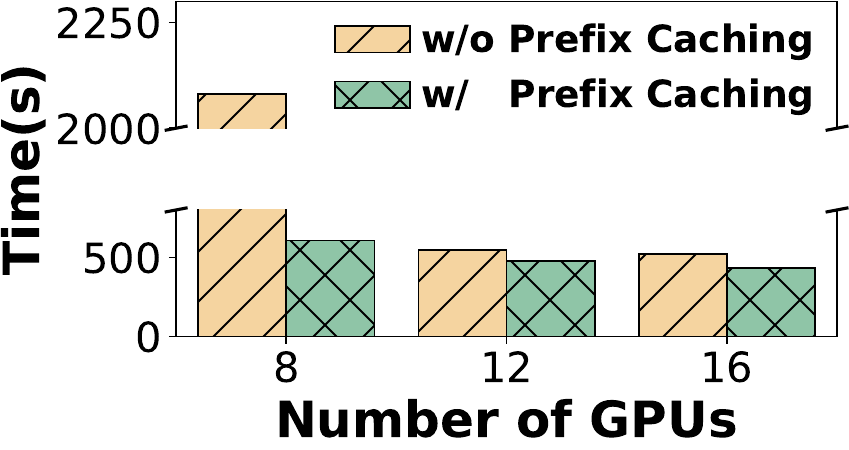}
    \caption{Impact of Prefill.}
    \label{subfig:prefill-token}
  \end{subfigure}\hfill
  \begin{subfigure}{0.24\textwidth}
    \centering
    \includegraphics[width=\linewidth]{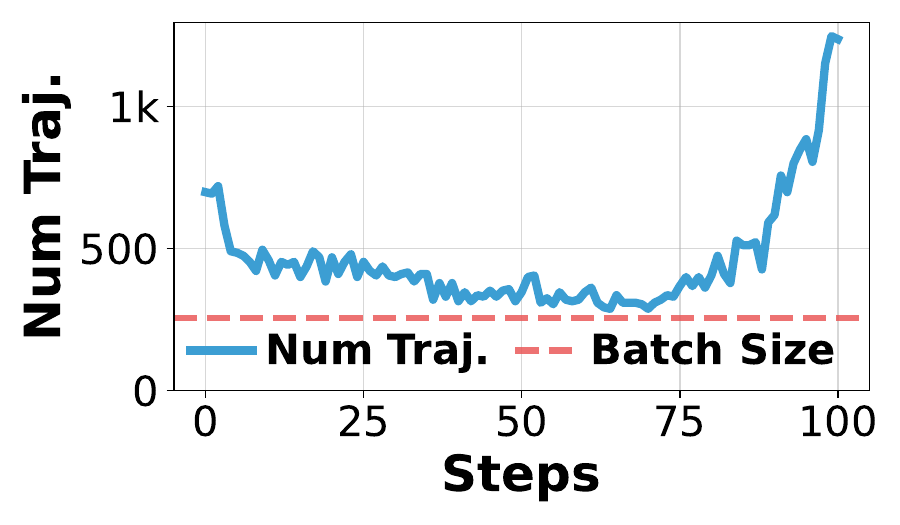}
    \caption{Need for Elasticity.}
    \label{subfig:elastic}
  \end{subfigure}\hfill
  \vspace{-10pt}
  \caption{Characterization of agentic RL: (a) The breakdown of end-to-end training time; (b) The long-tail distribution of rollout execution time; (c) The impact of prefill on rollouts; (d) The demand for resource elasticity.}
  \label{fig:workload-characterization}
  \vspace{-10pt}
\end{figure*}

The training time of agentic RL is dominated by the rollout stage, which accounts for over 70\% of total wall-clock time (\autoref{subfig:perf-breakdown}), with pronounced long-tail latency across rollout batches (\autoref{subfig:long-tail}).
Agentic rollouts further intensify GPU compute and memory demands.
First, multi-turn interactions require processing growing conversation histories at each turn, and the compute-intensive prefill phase benefits from prefix caching and resource scaling (\autoref{subfig:prefill-token}).
Second, agentic RL typically adopts redundant sampling to yield high-quality samples for gradient stability~\cite{DAPO,GRESO}, which exacerbates per-step resource contention.

These characteristics make rollout throughput highly responsive to resource scaling.
However, rollout resource demand varies significantly across training steps (\autoref{subfig:elastic}). The \emph{resource-fixed} systems~\cite{rollpacker,rollflash,areal,RollMux,sheng2024hybridflow,RhymeRL} that optimize within a \emph{static GPU budget} cannot adapt to this resource demand variation. An allocation sized for peak demand idles GPUs during light-load steps, while one sized for average demand creates contention during heavy-load steps.
This variation calls for resource elasticity. Existing \emph{resource-elastic} systems typically provision \textit{additional GPUs exclusively for rollouts} on demand using spot instances~\cite{rlboost} or serverless GPUs~\cite{thinkingmachines_tinker,openpipe_serverless_rl}.
However, these systems incur substantial allocation overhead during capacity churn (\autoref{fig:baseline_allocation_compare}), and the training performance is bounded by GPU availability, which can be scarce under cluster-wide contention~\cite{rlboost,RollMux}. 

A more natural source of elastic capacity for rollouts is the organization’s operational serving cluster. Our measurements show that serving GPUs average only 18.9\% compute utilization and 14.3\% memory utilization (\autoref{subfig:low-serving-util}), leaving substantial idle capacity for rollouts.
One intuitive approach is \emph{bidirectional autoscaling}, shrinking the serving cluster under low load and redirecting freed GPUs exclusively to rollouts.
However, reclaiming these GPUs for serving upon traffic bursts requires tens of seconds for model reloading and runtime initialization overhead~(\autoref{subfig:init-overhead}) that would violate serving Service Level Objectives~(SLOs).
Another option leverages \emph{GPU multiplexing} to co-locate rollout and serving workloads on the same GPU.
Yet existing GPU multiplexing systems fall short: those targeting homogeneous models~\cite{ConServe} cannot share KV cache (KVC) across heterogeneous layouts, while those designed for comparable SLO workloads~\cite{Prism,Aegaeon,duan2024muxserve} either cannot retain in-GPU rollout prefix cache for short-lived reuse or yield rollout KVC and compute cycles promptly under serving bursts.
The detailed analysis is in \S\ref{subsec:empirical-opportunity}.

In this paper, we explore \emph{cooperative elasticity}, where serving and rollout workloads cooperatively share \emph{already-deployed} GPUs within an organization to maximize rollout throughput while preserving serving SLOs. We present \SystemName{}, an agentic RL post-training framework to realize cooperative elasticity efficiently. While promising, repurposing serving GPUs for rollouts still poses two primary challenges: preserving serving SLOs under bursty traffic (\textbf{C1}) and heavy cross-cluster communication overhead (\textbf{C2}). First, co-locating heterogeneous serving and rollout LLMs must preserve serving SLOs under bursty traffic, despite contention for both GPU memory and compute.
To address this, we design a \emph{co-serving executor}~(\S\ref{subsec:local-executor}) with three modules:
(1) VMM-based cross-model KVC sharing for fast memory rebalancing across heterogeneous KVC layouts,
(2) Preemptive memory sharing that exploits the short-lived nature of rollout interactions to retain its prefix cache in GPU and reclaim it aggressively during serving bursts, mitigating memory contention,
and 
(3) Dual-SLO admission control leverages the priority asymmetry between rollouts, which can absorb second-level delays due to overlap among trajectories in the long tail, and serving, which requires millisecond-level SLOs, to prioritize serving through temporal sharing while opportunistically executing rollouts and avoiding severe compute interference.

Second, the serving and RL clusters may reside in different datacenters connected by bandwidth-limited links~(10--200\,Gbps Ethernet), so cross-cluster weight transfer can take tens of seconds to minutes~(\autoref{subfig:cluster-communication}), eroding the speedups from elastic rollouts. Existing communication systems~\cite{OmniReduce,ZEN,StellaTrain,sergeev2018horovod} assume fixed process groups with uniform sharding.
Cooperative elasticity breaks both assumptions: serving GPUs may join or leave across RL steps, while training and serving can adopt different parallelism strategies.
To mitigate this communication overhead, we introduce a \emph{cross-cluster weight transfer engine}~(\S\ref{subsec:weight-transfer}) that combines 
(1) A relay layer for asynchronous, fault-tolerant propagation,
(2) Shard-aware routing that automatically maps sharding rules across heterogeneous parallelism configurations,
and (3) Sparsity-aware compression that takes advantage of the lossless sparsity of RL weight deltas ({>}95\%, see~\autoref{fig:sparsity}) to transmit only non-zero elements, reducing transfer overhead to within 20\,seconds even under 20\,Gbps Ethernet~(\S\ref{subsec:eval-transfer-engine}).

Beyond these challenges, to fully capitalize on cooperative elasticity under time-varying serving load, \SystemName{} includes an \emph{elastic rollout scheduler} that dispatches rollouts across dedicated rollout GPUs and opportunistic serving GPUs~(\S\ref{subsec:global-scheduler}).
It combines (1) turn-wise concurrency-aware routing to offload excess trajectories when dedicated rollout GPUs saturate, and (2) cache-affinity placement to route each turn to the GPU holding its prefix KVC. It also provides fault-tolerant rerouting to migrate trajectories upon GPU stalls or failures.


We implement \SystemName{} atop ROLL~\cite{roll} and evaluate it with Qwen3-8B and Qwen3-32B~\cite{yang2025qwen3technicalreport} on agentic RL tasks~\cite{frozen_lake,ALFWorld} using 16--48 training GPUs and 16--64 serving GPUs (H800).
\SystemName{} improves average end-to-end training throughput by 1.3--3.3$\times$ over resource-fixed baselines~(ROLL~\cite{roll}, AReaL~\cite{areal}), and reduces rollout time by 1.2--1.5$\times$ relative to resource-elastic baselines~(RLBoost~\cite{rlboost}, $\lambda$RL~\cite{openpipe_serverless_rl,thinkingmachines_tinker}).
These gains come with nearly zero allocation overhead and no serving P99 SLO violations, validating \SystemName{} as a practical cooperative elasticity framework for agentic RL.

\section{Background and Motivation}
\label{sec:bg-workload-characterization}

\begin{figure}[tb]
  \centering
  \includegraphics[width=0.9\linewidth]{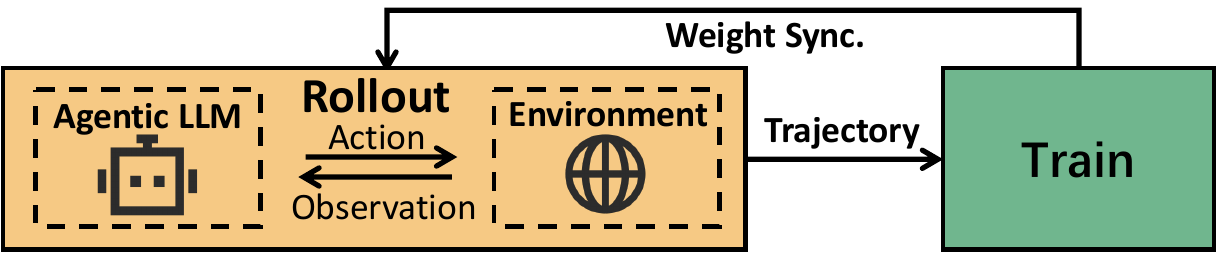}
  \vspace{-10pt}
  \caption{Illustration of Agentic RL Pipeline.}
  \label{fig:background}
  \vspace{-10pt}
\end{figure}

\subsection{Agentic RL Training}
\label{subsec:multi-task-agentic}
An agentic RL training pipeline alternates between \emph{rollout} and \emph{training}. In modern agentic RL, rollout is often organized using group-based algorithms~\cite{shao2024deepseekmath,DAPO,gigpo}. Taking GRPO as an example, the rollout stage launches $B_0$ \emph{environment groups}, where environments within the same group share the same initial state. For each group, the agent LLM (actor) interacts with the environment over multiple turns: at each turn it observes the current \emph{state}, samples an \emph{action} from its policy, and sends it back to the environment; the environment then transitions to a new state and returns feedback. The agent generates $G_0$ sampled responses per group, producing $G_0$ trajectories from the same starting point. When each trajectory receives the terminal signal, a reward worker evaluates it and assigns a scalar reward. In the training stage, the agent LLM updates its model weights using the collected trajectories and reward signals, and the updated weights are synchronized back to the rollout workers for the next step.

\subsection{Workload Characterization}
We perform workload characterization using Qwen3-8B and Qwen3-32B~\cite{yang2025qwen3technicalreport} within the agentic RL framework ROLL~\cite{roll}, configured with a maximum response length of 32k tokens, a batch size of 256, and a group size of 8 under the GRPO algorithm~\cite{shao2024deepseekmath}, on 16 H800 GPUs. We adopt synchronous RL training to profile the end-to-end training time. Specifically, we run the 8B and 32B LLMs on the agentic tasks FrozenLake and ALFWorld, respectively, for five consecutive steps.
 
\noindent\textbf{Dominant Rollout Overhead.} We report the performance breakdown in~\autoref{subfig:perf-breakdown}. The end-to-end training time consists of rollout, training, and weight synchronization overhead. The rollout stage accounts for over 70\% of total time and dominates end-to-end training. This motivates the tailored rollout optimizations to improve training efficiency. 



\noindent\textbf{Long-tail Rollouts.} \autoref{subfig:long-tail} presents the execution time distribution of trajectories during rollout for the 8B and 32B LLMs. We observe a pronounced long-tail pattern: most trajectories finish quickly, while a small fraction take much longer. In particular, the 75th percentile (P75) is at most 30\% of the end-to-end rollout time. This observation is consistent with prior studies~\cite{rollpacker,rollflash,letitflow}. Such long-tail phenomena can cause GPU underutilization, as many GPUs remain idle while waiting for straggler trajectories to complete.

\noindent\textbf{The Impact of Prefill.} Multi-turn agentic RL entails frequent prefills~\cite{rollart,letitflow}. In our workloads, prefill tokens account for 77\% (Qwen3-8B) and 86\% (Qwen3-32B) of total tokens, making prefill a dominant cost. As a result, prefix caching, which reuses KV cache (KVC) for shared prefixes across requests, is widely used in multi-turn rollouts~\cite{verl-agent,rollart,rollflash}. To quantify its impact, we vary the number of GPUs for Qwen3-8B and measure average rollout time with and without prefix caching in \autoref{subfig:prefill-token}. Under resource contention, prefix caching improves rollout throughput by up to $3.4\times$ (at 8 GPUs). As more GPUs are allocated, the rollout time reduces from 607\,s to 435\,s. This contrasts with single-turn RL, where scaling GPUs often yields limited benefit because long-tail samples are dominated by the bandwidth-bound decoding phase. In multi-turn agentic RL, scaling is more effective because of the compute-heavy prefill stage. Next, we analyze how rollout demand varies across steps to understand whether a static GPU allocation can be optimal.

\noindent\textbf{Need for Resource Elasticity.} In agentic RL training, agent--environment interaction often yields sparse rewards, and reward values within a group may exhibit low (or even zero) variance, weakening the learning signal. Hence, many algorithmic studies~\cite{DAPO,SPEEDRL,GRESO} propose redundant sampling: dynamically increasing $B_0$ to launch additional environment workers, and continuing rollouts until collecting $B_0$ groups with non-zero reward variance. We observe that many agentic RL jobs enable this feature in our internal cluster. 
\autoref{subfig:elastic} shows that when training a Qwen3-8B model with DAPO~\cite{DAPO}, the number of generated trajectories vary significantly across steps, with the maximum reaching $5.7\times$ the user specified  batch size.  
With a fixed GPU budget, redundant sampling increases memory pressure, and the high variability in rollout volume makes a static, resource-fixed configuration inefficient. Beyond redundant sampling, even standard GRPO with a fixed batch size benefits from resource elasticity: long-tail rollouts and compute-heavy prefills create variable per-step resource demand, and our end-to-end evaluation (\S\ref{subsec:end-to-end-evaluation}) confirms that resource elasticity improves average throughput by $1.31$--$1.46\times$ under GRPO. 
This motivates a \emph{resource-elastic} system that adjusts rollout GPUs over time to reduce contention and shorten step time.

\begin{figure*}[ht]
  \centering
   \begin{subfigure}{0.24\textwidth}
    \centering
    \includegraphics[width=\linewidth]{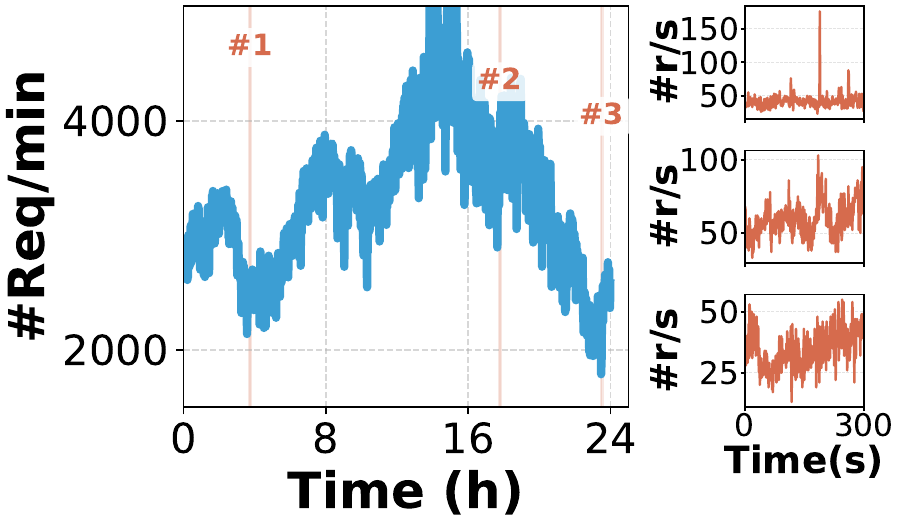}
    \caption{Serving Traffic.}
    \label{subfig:bursty-serving-traffic}
  \end{subfigure}\hfill
  \begin{subfigure}{0.24\textwidth}
    \centering
    \includegraphics[width=\linewidth]{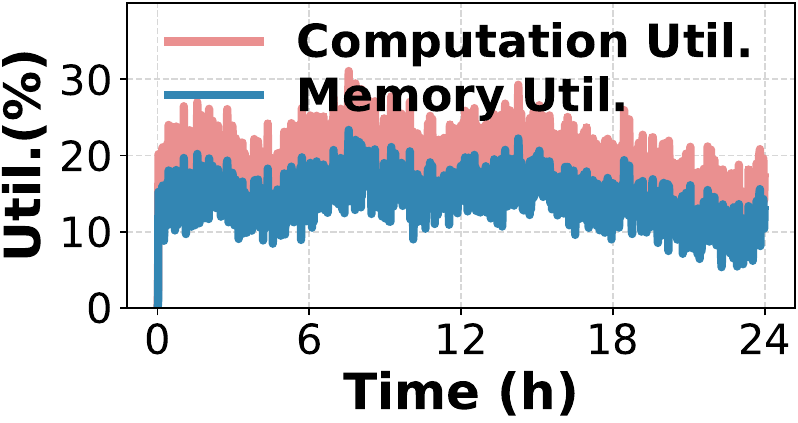}
    \caption{GPU Utilization.}
    \label{subfig:low-serving-util}
  \end{subfigure}\hfill
  \begin{subfigure}{0.24\textwidth}
    \centering
    \includegraphics[width=\linewidth]{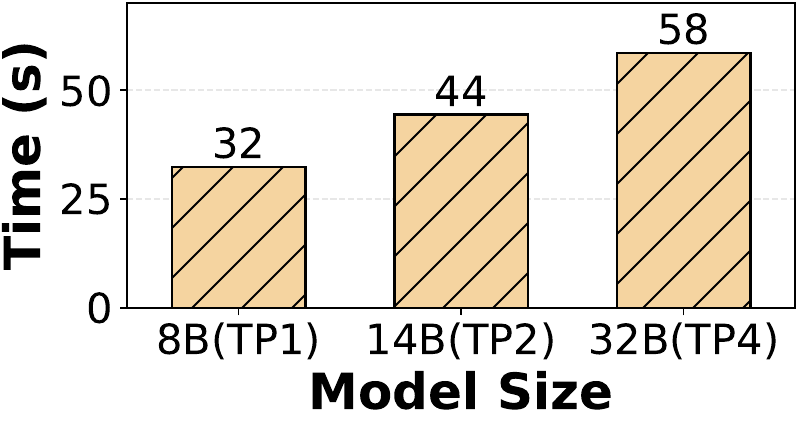}
    \caption{Allocation Overhead.}
    \label{subfig:init-overhead}
  \end{subfigure}\hfill
  \begin{subfigure}{0.24\textwidth}
    \centering
    \includegraphics[width=\linewidth]{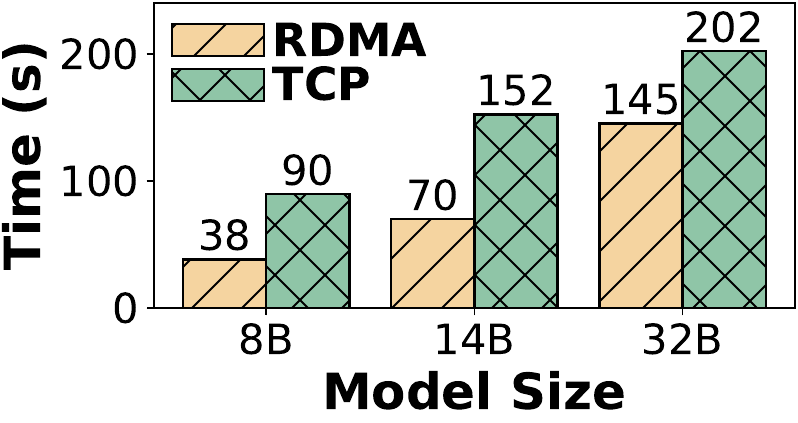}
    \caption{Comm Betw. Train\&Serve.}
    \label{subfig:cluster-communication}
  \end{subfigure}\hfill
\vspace{-5pt}
\caption{Characterization of serving clusters and workloads: (a) Fluctuating serving traffic; (b) Serving GPU underutilization; (c) High allocation overhead; (d) Substantial communication overhead.}
\label{fig:serving-characterization}
\vspace{-5pt}
\end{figure*}

\begin{figure}[tb]
  \centering
    \centering
    \includegraphics[width=0.9\linewidth]{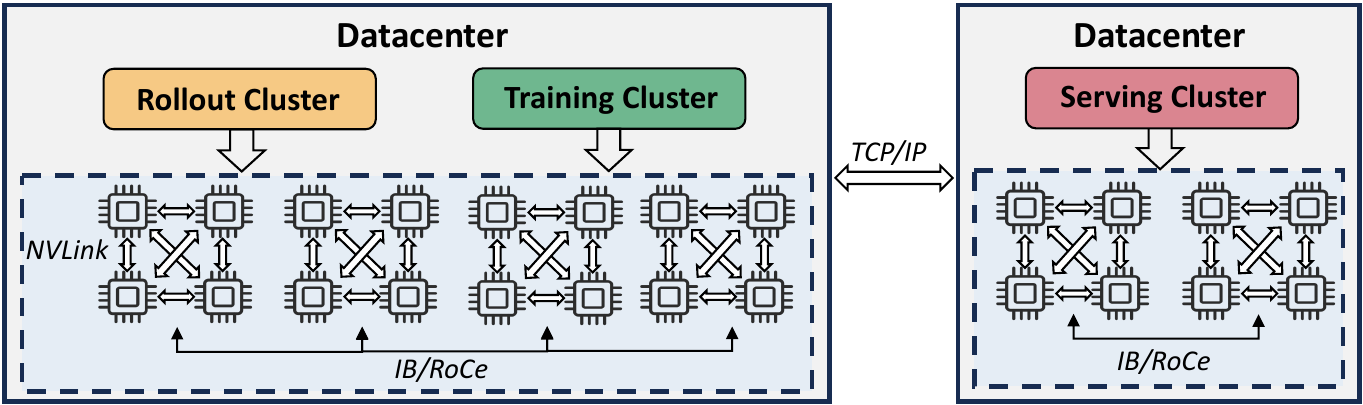}
    \vspace{-10pt}
  \caption{Scheme of Datacenter Infrastructure.}
  \label{fig:infra}
  \vspace{-10pt}
\end{figure}

\subsection{Limitations of Existing Solutions}
\noindent\textbf{Resource-fixed RL systems struggle to balance utilization and training time.}
Many existing RL training systems~\cite{verl-agent,roll,RhymeRL,dist-aware-sd,RollMux,StreamRL,MiMo,zhong2024rlhfuseefficientrlhftraining,skyrl-agent,areal,Laminar,rollflash,asyncflow,rollpacker} provision a fixed number of GPUs for rollouts throughout training. This static allocation cannot align the actual rollout workload demand: rollout latency exhibits heavy tails, so overprovisioning leaves GPUs underutilized due to long-tail trajectories, while underprovisioning increases contention for compute and memory, inflating rollout time. Moreover, as \autoref{subfig:elastic} shows, rollout demand varies across steps, making any single fixed allocation suboptimal over time.

\noindent\textbf{Elastic RL systems suffer from high allocation overhead.}
Existing resource-elastic designs rely on preemptive GPUs (e.g., spot instances~\cite{rlboost}, serverless GPUs~\cite{thinkingmachines_tinker,openpipe_serverless_rl}) to scale rollout capacity on demand. Because these GPUs lie outside the steady-state deployment, each provisioning event requires model loading and runtime initialization, which can take tens of seconds (\autoref{subfig:init-overhead}). Spot preemption and serverless lease expiration further trigger repeated teardown-and-reinitialize cycles, turning allocation overhead into a persistent throughput tax that can consume over 20\% of total training time (\autoref{fig:micro_benchmarks_e2e}). Cluster-wide resource contention exacerbates this problem: as spot capacity becomes scarce~\cite{rlboost} or serverless pools fail to secure GPUs under concurrent RL jobs, the system falls back to fewer GPUs with more frequent reallocation attempts, further inflating allocation overhead and undermining the promised elasticity.

\section{Opportunities and Challenges}
\label{sec:opportunity-challenge}
We first describe the infrastructure setting, then quantify the harvestable serving capacity, explain why strawman approaches fail to exploit it, and highlight the key challenges.

\subsection{Infrastructure for RL and Serving}
\label{subsec:infra}
Many AI organizations operate separate RL and serving infrastructures (\autoref{fig:infra}). We categorize GPUs into three clusters: a \emph{training cluster}, a \emph{rollout cluster}, and a \emph{serving cluster}. Training and rollout clusters are co-located and communicate via high-speed interconnects (e.g., NVLink, InfiniBand). Serving clusters may reside in separate datacenters for operational isolation~\cite{jouppi2023tpuv4,megascale,Aegaeon}, connected by bandwidth-limited links (10--200\,Gbps Ethernet)~\cite{RollMux,rollart}.
RL infrastructure and LLM serving infrastructure are operated by different teams within the same organization, each managing large-scale GPUs. Since rollout and serving are both inference workloads, both teams often collaborate on optimizing a shared LLM inference engine, a pattern also observed in open-source communities~\cite{sglang}. This organizational proximity makes it natural to consider repurposing serving resources for rollouts when rollout demand spikes.

\subsection{Harvestable Serving Capacity}
\label{subsec:empirical-opportunity}
We next quantify how much serving capacity is empirically available for rollouts and explain why strawman approaches fail to exploit it.

\noindent\textbf{Fluctuating Serving Traffic Leaves GPU Underutilized.} Production LLM serving workloads exhibit fluctuating request rates~\cite{BurstGPT,Prism,kvcachecachewild,llm-multitasking,qlm2024patke}. \autoref{subfig:bursty-serving-traffic} plots a 24-hour Microsoft trace~\cite{stojkovic2025dynamollm} at minute granularity alongside three zoomed-in 5-minute windows at per-second granularity. At the minute level, the peak rate reaches $1.7\times$ the 24-hour average. At the second level, burstiness is far more pronounced: per-second peaks reach 4.22$\times$, 1.58$\times$, and 1.73$\times$ their respective window averages, consistent with second-level spikes reported by BurstGPT~\cite{BurstGPT}. To absorb such spikes, providers often statically overprovision for peak demand~\cite{distserve}, resulting in substantial GPU underutilization. We quantify this by replaying a 24-hour production trace from~\cite{qlm2024patke} (preserving original prompt lengths, response lengths, and arrival process) on Qwen3-8B with 8 H800 GPUs. \autoref{subfig:low-serving-util} shows the GPU utilization sampled at 1-second intervals and smoothed with a one-minute moving average. On average, GPUs reach only 18.9\% SM utilization and 14.3\% HBM utilization, confirming that peak-provisioned GPUs remain significantly underutilized for much of the day. Beyond per-GPU underutilization, serving clusters often comprise thousands of GPUs~\cite{Aegaeon}, so even modest slack per GPU aggregates into a large pool of idle cycles and memory. A natural question is whether this spare capacity can be harvested for rollouts.

\noindent\textbf{Strawman Approaches Cannot Safely Harvest the Capacity.}
A natural first attempt is \emph{bidirectional autoscaling}~\cite{lambdascale,BLITZSCALE}, which shrinks the serving cluster during low load and redirects freed GPUs to rollouts. However, bidirectional autoscaling is fundamentally limited: reclaiming GPUs from rollouts back to serving requires evicting in-flight rollouts and reloading models, taking tens of seconds (\autoref{subfig:init-overhead}) and far exceeding typical SLO budgets. Because serving traffic is bursty at second-level granularity, frequent mode switching triggers repeated initialization overhead that erodes throughput gains.
An alternative is \emph{GPU multiplexing}~\cite{ConServe,Prism,Aegaeon,duan2024muxserve}, which co-locates both workloads on the same GPU to avoid repeated initialization. However, co-locating heterogeneous models with asymmetric SLO requirements introduces challenges in memory sharing and compute interference that existing systems do not address~(detailed in \S\ref{subsec:challenge}).
Our evaluation~(\S\ref{subsec:end-to-end-evaluation}) confirms that neither approach can safely harvest serving slack: both autoscaling~\cite{fu2024serverlessllm} and multiplexing~\cite{Prism} violate serving SLOs and degrade rollout throughput.

These failures motivate a different approach. Rather than GPU-level switching or treating rollouts as generic co-located inference, we explore \emph{cooperative elasticity}: keeping both serving and rollout models resident on each GPU and dynamically sharing resources, exploiting properties of agentic RL rollouts that make such sharing safe and efficient. While this can exploit abundant serving capacity with minimal allocation overhead, it raises two key challenges.

\subsection{Challenges in Cooperative Elasticity}
\label{subsec:challenge}
\noindent\textbf{C1: Preserving Serving SLOs under Bursty Traffic.} Cooperative elasticity keeps heterogeneous serving and rollout models deployed on the same GPUs, but preserving serving SLOs under bursty traffic requires accounting for both memory and compute contention in this deployment.

\begin{itemize}[leftmargin=*,noitemsep,topsep=2pt]
  \item \textbf{Inefficient memory sharing.} Serving and rollout models are often heterogeneous with incompatible KVC layouts. Static reservation of per-model KVC pools~\cite{vllm,sglang,ConServe} requires runtime reinitialization, which can take tens of seconds (\autoref{subfig:init-overhead}). Heterogeneous colocation engines~\cite{Prism,Aegaeon} typically omit in-GPU prefix caching or do not account for KVC contention between prefix cache entries and active request KVC across co-located models.
  Since rollouts benefit significantly from prefix caching, this contention        
  competes with serving KVC for limited HBM and inflates latency under traffic     
  spikes, triggering SLO violations.

  \item \textbf{Severe compute interference.}
  LLM serving systems enforce latency SLOs on \emph{time-to-first-token} (TTFT) and \emph{time-per-output-token} (TPOT). Several systems disaggregate prefill and decoding onto separate GPU pools (PD disaggregation) to stabilize TTFT/TPOT under bursty traffic~\cite{patel2023splitwise,distserve}. Unlike standard multi-tenant serving, where co-located workloads share comparable SLO requirements, rollout inference has fundamentally asymmetric characteristics: rollouts run a \emph{different} LLM with much looser latency targets, generate long multi-turn sequences that sustain GPU occupancy for seconds, and perform both prefill and decode on the same GPU (PD co-location) even when serving deployment uses PD disaggregation. Standard request scheduling cannot bound the compute interference: a single rollout prefill chunk can delay serving decodes, and sustained rollout batches can starve serving prefills (analyzed in \S\ref{subsec:analysis-co-serving-executor}).
\end{itemize}

\noindent\textbf{C2: Heavy Cross-cluster Communication Overhead.}
In many deployments, RL training and online serving are provisioned as separate GPU clusters, and weight updates must traverse cross-cluster links via a flexible transfer engine. To quantify this overhead, we measure the end-to-end time to transfer LLM parameters of varying sizes from a GPU node in the RL cluster to a GPU node in the serving cluster using \texttt{Mooncake Store}~\cite{qin2024mooncake}\footnote{The weight transfer adopted here is a \emph{batch} baseline in \S\ref{subsec:eval-transfer-engine}.}, over TCP (200\,Gbps Ethernet) and RDMA (400\,Gbps InfiniBand), shown in \autoref{subfig:cluster-communication}. Even with InfiniBand (which is uncommon across datacenters), it can take up to 145\,s and grow quickly with model size, becoming a bottleneck for frequent weight synchronization.
\begin{figure}[tb]
  \centering
    \centering
    \includegraphics[width=\linewidth]{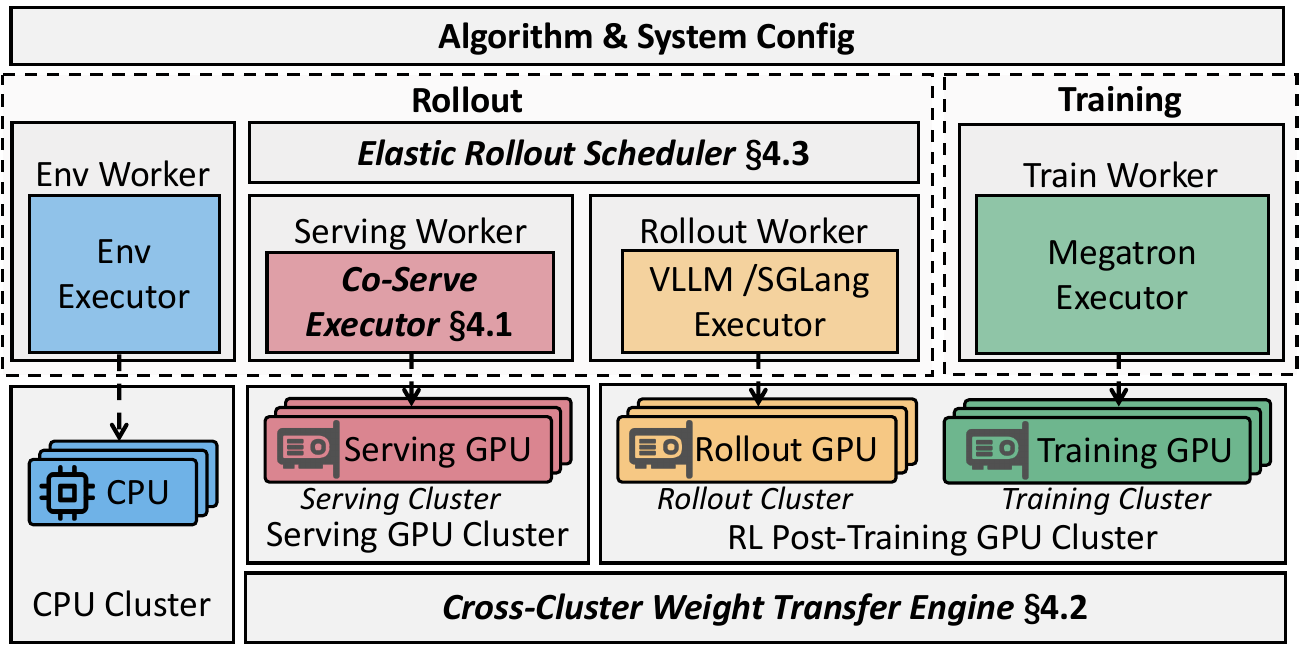}
    \vspace{-15pt}
  \caption{System Architecture of \SystemName{}.}
  \vspace{-10pt}
  \label{fig:overview}
\end{figure}

\section{System Design}
\label{sec:overview}
\noindent\textbf{System Overview.} To address the above challenges, we introduce \SystemName{}, the architecture of which is illustrated in \autoref{fig:overview}. It includes three novel components: (1) an SLO-safe \emph{co-serving executor} that co-locates \emph{heterogeneous} serving and rollout LLMs on the same GPUs, dynamically sharing memory and compute while preserving serving SLOs (\S\ref{subsec:local-executor}); (2) a cross-cluster \emph{weight transfer engine} that asynchronously synchronizes model weights from the training cluster to the serving cluster with sharding and sparsity awareness (\S\ref{subsec:weight-transfer}); and (3) an elastic \emph{rollout scheduler} that dispatches rollouts across rollout and serving GPUs to realize the cooperative elasticity and achieve end-to-end training throughput improvement under varying serving and rollout load (\S\ref{subsec:global-scheduler}).

\noindent\textbf{Serving Cluster Setup.}
\SystemName{} supports diverse serving deployments, including PD disaggregation vs.\ co-location and autoscaled vs.\ statically provisioned clusters. In our evaluation, we use PD disaggregation with a statically provisioned serving cluster and enforce P99 tail-latency SLOs on TTFT and TPOT. To avoid the initialization overhead of loading rollout models on demand, we pre-deploy commonly used rollout models alongside the heterogeneous serving LLMs in the serving cluster and expose endpoints for weight transfer and response generation. Rollout workers remain deactivated when idle and are activated during the rollout stage. We detail the concrete model pairing and cluster setup in \S\ref{ssec:eval_setup}.

\noindent\textbf{System Workflow.} The user specifies the RL resource request $N_{\text{rl}}$ and an upper bound on the number of serving GPUs that can be borrowed $N_{\text{serving}}$.
Upon job submission, the RL cluster reserves $N_{\text{rl}}$ GPUs, and the serving cluster selects up to $N_{\text{serving}}$ GPUs with the lowest recent KVC usage over a fixed window (e.g., 1 hour).
To avoid contention among concurrent RL jobs, \SystemName{} assigns each selected serving GPU to at most one RL job for rollouts.
During this setup, \SystemName{} initializes the RL runtime on the reserved RL GPUs and launches rollout workers accordingly. On selected serving GPUs, \SystemName{} activates the rollout runtime into GPU memory.

Each RL step follows a typical RL pipeline. First, \SystemName{} initiates the rollout stage. Each trajectory involves multi-turn environment interactions, and the elastic rollout scheduler dispatches each turn to either dedicated rollout GPUs or borrowed serving GPUs to improve rollout throughput while enforcing serving SLOs. Second, on each serving GPU, the co-serving executor dynamically shares memory and compute between the heterogeneous serving and rollout LLMs, enforcing serving TTFT and TPOT SLOs while harvesting otherwise idle GPU cycles for rollouts. If serving load causes rollout stalls, the scheduler reroutes subsequent trajectories to underutilized GPUs. Third, the generated trajectories are fed into the training stage. Once training produces updated weights, \SystemName{} invokes the cross-cluster weight transfer engine, which asynchronously pushes weight updates to serving GPUs while the next step's intra-cluster synchronization and rollout execution proceed in parallel (\S\ref{subsec:weight-transfer}).

\subsection{SLO-Safe Co-Serving Executor}
\label{subsec:local-executor}

The co-serving executor runs rollouts opportunistically on serving GPUs while preserving serving TTFT and TPOT SLOs. This requires managing two sources of contention: \emph{GPU memory} (dominated by KVC) and \emph{GPU compute} (which directly affects token-level latency). 
As discussed in \S\ref{subsec:challenge}, existing multiplexing systems cannot simultaneously meet these requirements.
We address these challenges through three techniques, enforcing two principles: \emph{serving-first memory} (serving always has priority on KVC capacity) and \emph{serving-first compute} (serving always has priority on GPU compute).

\noindent\textbf{VMM-based Cross-Model KV Cache Memory Sharing.} 
\yuheng{Heterogeneous serving and rollout models have incompatible KVC layouts that preclude dynamic memory rebalancing in mainstream engines~(\S\ref{subsec:challenge}).}
We leverage CUDA Virtual Memory Management (VMM) to enable fast, flexible KVC rebalancing across heterogeneous models. We decouple \emph{virtual} KV address spaces from \emph{physical} GPU pages: each model reserves a contiguous virtual KV address space that preserves its attention-kernel indexing, while all models share a global physical page allocator that maps and unmaps pages on demand. When serving load increases, we unmap physical pages from rollout’s virtual address space and remap them into serving’s virtual address space at page granularity (typically 2MB). This enables cross-model memory rebalancing without modifying KVC layouts. We keep a lightweight runtime context warm on serving GPUs for fast rollout model (re-)activation. Activating Qwen3-32B completes within 5\,s via PCIe/NVLink weight loading, avoiding the tens-of-seconds overhead of add-capacity elasticity.

\noindent\textbf{Preemptive Memory Sharing Policy.} Prefix caching is critical for rollout performance (\autoref{subfig:prefill-token}), but existing multiplexing systems~\cite{Prism,Aegaeon} 
\yuheng{either offload completed request KVC to CPU memory or do not manage prefix cache contention across co-located models.} 
However, rollout environment interactions are short-lived: consecutive interactions typically occur within 10\,s~\cite{rollart}, concentrating prefix reuse in a brief window. CPU-GPU transfer latency makes offloading ineffective for such short intervals. We instead keep rollout prefix cache resident in GPU to exploit short-term reuse, but this creates memory contention with serving that prior systems do not address. We note that prior KVC engines~\cite{qin2024mooncake,cheng2025lmcache} also use host memory offloading, which is ineffective for rollouts because weights are updated frequently, invalidating cached entries while increasing high CPU memory pressure.

We address this tension through \emph{preemptive memory sharing}: we keep rollout prefix cache in GPU under typical load, but aggressively reclaim it during serving bursts, relying on the short interaction window to recapture most reuse before entries expire. At the beginning of each RL step, the elastic rollout scheduler derives a per-GPU rollout KVC budget from the serving model's recent memory usage and reserves a fixed KVC headroom $H$ (e.g., 20\% of total GPU memory) for serving. This iteration-level budget stabilizes rollout memory usage under typical serving load, but it cannot react to sudden serving bursts.

The memory sharing policy proceeds in three steps. (1) \emph{Burst trigger:} The co-serving executor continuously monitors serving KVC usage. When serving starts to consume the reserved headroom (i.e., serving KVC usage crosses a high-watermark within $H$), the executor enters a \emph{pressure} state. (2) \emph{Emergency cut:} In the pressure state, the executor immediately shrinks the rollout KVC budget by a fixed factor (2$\times$) and returns the freed physical pages to the shared allocator. It reclaims rollout KVC pages at request granularity, aborts the affected rollout requests, and notifies the rollout scheduler (\S\ref{subsec:global-scheduler}) to reroute the affected trajectories to other underutilized GPUs. This one-time aggressive cut avoids repeated fine-grained reallocations and reduces allocation/reclamation churn during load bursts. (3) \emph{Freeze:} To prevent oscillation between reclaiming and regrowing rollout KVC under bursty traffic, the executor does not increase the rollout budget until the next RL step, when the rollout scheduler recomputes budgets using updated serving statistics. This conservative choice preserves serving SLOs, while the rollout scheduler absorbs the transient capacity loss by shifting subsequent rollout steps to underutilized GPUs, including dedicated rollout GPUs, as rollouts progress. We attach a short lease (e.g., 10\,s) to each rollout KVC page and reclaim pages when the lease expires, bounding HBM consumption while capturing most prefix reuse within the typical environment interaction window.

\noindent\textbf{Dual-SLO Admission Controller.} Co-serving causes compute interference between rollout and serving under spatial co-location (analyzed in \S\ref{subsec:analysis-co-serving-executor}). Unlike serving requests that require millisecond-level TTFT (e.g., 100\,ms) and TPOT (e.g., 50\,ms) SLOs, 
\yuheng{
individual rollout stages exhibit pronounced long-tail latency, with P75 around 55--145s and tails exceeding 400s~(\autoref{subfig:long-tail}). 
This long-tail overlap, together with trajectory rerouting by the elastic rollout scheduler (\S\ref{subsec:global-scheduler}), allows \SystemName{} to absorb second-level rollout delays without materially hurting the performance.
}
We exploit this asymmetry through \emph{temporal sharing} with serving-first admission control: only one workload executes on the GPU at a time while both reside in GPU memory, serving tokens are always prioritized, and rollout tokens are admitted only when sufficient SLO slack exists. When serving bursts arrive, we immediately yield compute---stalled rollouts simply migrate to underutilized GPUs without permanent progress loss. 

We pre-profile runtime costs for both serving and rollout execution and use them to make online admission decisions. For prefill, we profile the latency of monolithic and chunked prefill as a function of prompt length, denoted by $\hat{T}_{\mathrm{prf}}(L, m)$ where $m \in \{\textsf{mono}, \textsf{chunk}\}$. For decode, we profile the per-step latency as a function of batch size, denoted by $\hat{T}_{\mathrm{dec}}(b)$. These profiles allow the executor to estimate, at each scheduling tick, the remaining slack under the serving TTFT and TPOT SLOs, and to admit rollout tokens only when sufficient slack is available. Although the serving stack uses PD disaggregation, we perform PD colocation for rollouts to maximize serving resource utilization. We further enable chunked prefill for rollouts (e.g., chunk size 512 tokens), which bounds the runtime of each rollout step and avoids head-of-line blocking for serving. Next, we describe the computation of SLO slack on prefiller and decoder instances.

On \textbf{prefiller} instances, we first compute the TTFT slack of pending serving requests at each scheduling tick. For a serving request $r$, let $t_{\mathrm{now}}$ be the current time and $t_r^{\mathrm{arr}}$ be its arrival time. Let $B_{\mathrm{TTFT}}$ be the configured TTFT SLO budget. Given prompt length $L_r$ and the serving prefill mode $m$, the remaining TTFT slack is
\begin{equation}
S^{\mathrm{prf}}_r
=
\left(t_r^{\mathrm{arr}} + B_{\mathrm{TTFT}}\right)
-
t_{\mathrm{now}}
-
\hat{T}_{\mathrm{prf}}(L_r, m).
\end{equation}
We conservatively use the minimum slack among queued prefills as the TTFT slack for rollouts, \(S^{\mathrm{prf}}_{\min} = \min_{r \in \mathcal{Q}_{\mathrm{prf}}} S^{\mathrm{prf}}_r\).

On \textbf{decoder} instances, for a serving request $r$, let $t_r^{\mathrm{last}}$ be the time it produced its most recent token.
Let $B_{\mathrm{TPOT}}$ be the configured TPOT budget, and let $b$ be the current serving decode batch size.
The remaining TPOT slack is
\begin{equation}
S^{\mathrm{dec}}_r
=
\left(t_r^{\mathrm{last}} + B_{\mathrm{TPOT}}\right)
-
t_{\mathrm{now}}
-
\hat{T}_{\mathrm{dec}}(b).
\end{equation}

We again use the minimum slack among active decodes, \(S^{\mathrm{dec}}_{\min} = \min_{r \in \mathcal{Q}_{\mathrm{dec}}} S^{\mathrm{dec}}_r\). We admit rollout token generation only when two conditions hold. First, the serving workload has sufficient TTFT and TPOT slack to accommodate the additional compute for rollout tokens. Second, allocating the corresponding KVC pages for rollout does not reduce the serving model's available KVC capacity below a reserved headroom. If rollout prefill or decoding makes no progress for a fixed timeout (e.g., 2 seconds), the co-serving executor reports a stall to the global scheduler and drops this trajectory, allowing the scheduler to reroute it elsewhere.

\subsection{Cross-Cluster Weight Transfer Engine}
\label{subsec:weight-transfer} 
\begin{figure}[tb]
  \centering
  \begin{subfigure}[b]{0.48\linewidth}
    \centering
    \includegraphics[width=\linewidth]{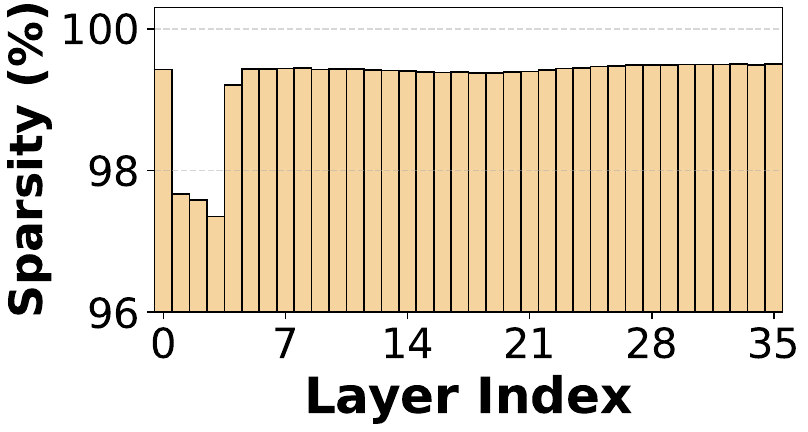}
    \caption{FrozenLake-8B.}
    \label{fig:sparsity_8B}
  \end{subfigure}\hfill
  \begin{subfigure}[b]{0.48\linewidth}
    \centering
    \includegraphics[width=\linewidth]{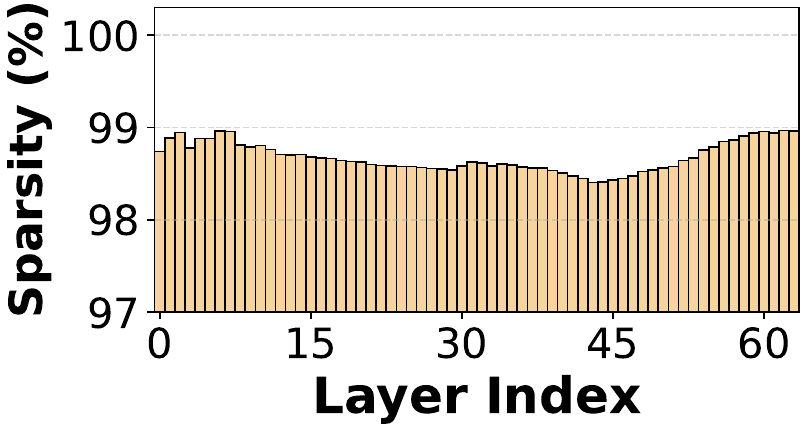}
    \caption{ALFWorld-32B.}
    \label{fig:sparsity_32B}
  \end{subfigure}
  \vspace{-10pt}
  \caption{Layer-wise sparsity ratio at 10th step.}
  \label{fig:sparsity}
  \vspace{-20pt}
\end{figure}

Cross-cluster communication overhead can bottleneck cooperative elasticity (\S\ref{sec:opportunity-challenge}). Existing communication systems~\cite{OmniReduce,ZEN,StellaTrain,sergeev2018horovod} optimize conventional model training under low-bandwidth links by adjusting training configurations or exploiting gradient sparsity and quantization. However, they assume collective communication over fixed process groups with uniform sharding, typically under data parallelism. Cooperative elasticity introduces two requirements absent in these settings: (1) dynamic GPU membership, as serving GPUs join or leave across RL steps, precludes fixed collectives; and (2) heterogeneous parallelism strategies between training and serving require automatic shard mapping. Additionally, we observe that RL post-training \emph{weight deltas} are $>$95\% sparse (\autoref{fig:sparsity}), a property uncommon in conventional LLM training that enables lossless compressed delta transfer for \emph{cross}-cluster weight synchronization. We address these through three techniques.

\noindent\textbf{Asynchronous Weight Transfer.} Resource elasticity makes the set of participating serving GPUs \emph{dynamic}: serving GPUs may join or leave between RL steps due to changing serving load. This precludes fixed collective groups and requires fault-tolerant, point-to-point transfer that gracefully handles membership changes. Following RollArt~\cite{rollart}, we build the transfer engine on \texttt{Mooncake Store} as a relay layer that decouples training and serving: training workers push weights in fixed-size buckets (e.g., 64\,MB) to relay workers asynchronously, and serving workers pull in larger batches (e.g., 1\,GB) on demand without coordinating with training or other serving workers. This avoids establishing fixed communication groups and makes transfer robust to membership changes. We overlap cross-cluster transfer with NCCL-based intra-cluster synchronization so that rollout workers can resume without waiting for cross-cluster transfer to finish.

\noindent\textbf{Shard-aware Weight Transfer.} Training and serving clusters adopt heterogeneous parallelism strategies (e.g., training with TP8$\times$PP2 and serving with TP4), requiring automatic shard mapping across configurations. Naive approaches require manual resharding or full model aggregation before transfer. \SystemName{} automatically infers each parameter's sharding rule by identifying the sharded dimension from the module type and parameter shape, computing per-rank slice ranges, and encoding this metadata in the Mooncake object key. On the training side, each device pushes its local shard asynchronously rather than first all-gathering the full model. To avoid redundant sends, each data-parallel rank transmits a mutually exclusive set of shards, parallelizing transfers and improving link utilization. On the serving side, each rank derives which buckets to pull based on the encoded metadata, fetching only its needed shards rather than a full model replica. This substantially reduces transfer volume and makes transfer transparent to the RL algorithm. We support tensor parallelism (TP) and pipeline parallelism (PP).

\noindent\textbf{Sparsity-aware Weight Transfer.} 
Transfer time scales with data volume, so naively synchronizing full weights becomes prohibitive as model size grows. 
Lossless gradient sparsity is rare in LLM training~\cite{muhamed2024grass,sarfi2025communicationefficientllmpretraining,yang2025sparsegradientcompressionfinetuning}, but we observe that RL post-training produces a natural source of lossless sparsity: the \emph{weight differentials} $\Delta W_t = W_t - W_{t-1}$ between consecutive steps are highly sparse, because RL algorithms employ gradient-stabilization techniques (e.g., reference models, KL penalties, and conservative update rules)~\cite{DAPO,shao2024deepseekmath} that constrain policy drift. 
We define \emph{sparsity ratio} as the fraction of zero elements in $\Delta W_t$. \autoref{fig:sparsity} reports the layer-wise sparsity for Qwen3-8B and Qwen3-32B at the $10^{th}$ RL step. Across layers, more than 95\% of elements in $\Delta W_t$ are zero. \S\ref{subsec:eval-transfer-engine} confirms that this high sparsity persists across RL steps. The engine can therefore ship sparse deltas rather than full replicas.

To exploit this sparsity efficiently, we compress $\Delta W_t$ in COO (coordinate) format for transfer and keep the previous-step weights $W_{t-1}$ resident on local devices. At each update, we reconstruct the current weights by applying $\Delta W_t$ to $W_{t-1}$. This additive update introduces at most $\sim$1\,s of compute overhead, which is small relative to the cross-cluster transfer time (up to minutes). To further reduce overhead, we shard the COO tensors according to the chosen parallelism strategy so that each device applies only its local shard, avoiding sparse-to-dense materialization and redundant additions.

\subsection{Elastic Rollout Scheduler}
\label{subsec:global-scheduler}
The rollout scheduler harnesses cooperative elasticity to orchestrate trajectory generation across rollout and serving GPUs. However, rollout workers on serving GPUs expose a response generation endpoint, and their data plane differs from that of rollout workers running on dedicated rollout GPUs. We therefore introduce a unified rollout proxy to allow the scheduler to manage heterogeneous rollout workers sharing a common interface for response generation and per-trajectory metric collection. Next, we describe how the scheduler decides how many trajectories to offload to serving GPUs and how it places each trajectory.

\noindent\textbf{Turn-wise Concurrency-aware Routing.}
The GPU compute and memory jointly bound the number of trajectories that can execute efficiently on the dedicated rollout GPUs. We perform offline profile and cap rollout concurrency at a workload-dependent threshold (e.g., 16 per GPU for Qwen3 models with 32K context length) to avoid KVC pressure and excessive scheduling overhead. When the instantaneous rollout demand exceeds this limit, the scheduler offloads the excess trajectories to available serving GPUs. Since multi-turn agentic RL alternates environment interaction with LLM generation, \SystemName{} schedules rollouts at \emph{turn} granularity rather than pinning each trajectory to a single GPU. This fine-grained control allows subsequent turns of a trajectory to spill over to serving GPUs under contention, and to migrate back to rollout GPUs once capacity becomes available.

\noindent\textbf{Cache-Affinity Placement.}
To maximize the benefit of prefix caching, the rollout scheduler employs a cache-affinity placement policy across both dedicated rollout GPUs and opportunistic serving GPUs. For each trajectory, the scheduler records the rollout worker that served its previous turn, which typically retains the trajectory’s prefix KVC. For each new turn, the scheduler first routes the request to the cache-affine worker if it has available capacity on a rollout GPU or, for a serving GPU, if admitting the request would not violate serving SLOs. If the cache-affine worker is unavailable, the scheduler falls back to a load-aware policy by dispatching the request to the least-loaded rollout GPU when possible; otherwise, it dispatches to the least-loaded eligible serving GPU. If neither pool has capacity, the request is queued until resources become available.

\noindent\textbf{Fault Tolerance and Recovery.}
The rollout scheduler uses a heartbeat mechanism to monitor the liveness of each serving worker. When it receives an execution stall signal from the co-serving executor (\S\ref{subsec:local-executor}) or detects a failure via health checks, the scheduler promptly reroutes the affected trajectories to other available GPUs, enabling fast recovery.

\noindent\textbf{Extensions to Other Deployment Settings.}
\SystemName{} can generalize to two other LLM serving deployments. \emph{PD colocation.} When prefill and decode are co-located on the same serving instance, the executor derives per-tick compute slack for rollout admission as the minimum one between TTFT and TPOT slack. \emph{Autoscaling.} This can leave a fraction of GPUs idle under low traffic. \SystemName{} can run rollouts on these idle GPUs and leverage the fast model swap mechanism in \S\ref{subsec:local-executor} to utilize GPU compute and memory without contending with serving traffic.
Due to the cost of large-scale evaluation across multiple production configurations, we focus on the mainstream PD-disaggregated deployment used in industry.

\section{Implementation}
\label{sec:implementation}
We implement $\sim$5k lines of Python atop agentic training framework ROLL~\cite{roll} and serving framework vLLM~\cite{vllm}. 

\noindent\textbf{Agentic RL training.} The rollout scheduler and the push side of the weight transfer engine are implemented inside ROLL. \SystemName{} uses Megatron-LM~\cite{shoeybi2019megatron} for training, vLLM for rollout/serving with request migration, and ROLL's native environment runtime to manage environments.

\noindent\textbf{Serving engine.} The pull side of the transfer engine and the co-serving executor are built atop vLLM 0.10.0~\cite{vllm}. We deploy rollout models in the serving cluster using \texttt{vllm serve}, exposing endpoints for weight transfer and response generation. When idle, rollout workers remain deactivated (not GPU-resident) and consume at most 2\,GB GPU memory; they are activated during the rollout stage. We implement a Ray-based~\cite{ray} load-balancing policy to route serving requests.

\noindent\textbf{Relay worker.} We use Mooncake v0.3.8 in the relay worker, allowing ROLL to publish updated weights and vLLM to pull them. We extend Mooncake with shard awareness and sparsity awareness to reduce communication overhead.

\noindent\textbf{Environment runtime.} Environments run in CPU-only containers on a separate Kubernetes cluster, communicating with rollout workers via K8S API calls. This isolates environment execution from GPU workloads. \SystemName{}'s design is orthogonal to environment placement and extends to GPU-accelerated environments.

\section{Performance Evaluation}
\label{ssec:eval_setup}
\noindent\textbf{Models and Training Configurations.} We use Qwen3-8B/32k for FrozenLake~\cite{frozen_lake} and Qwen3-32B/32k for ALFWorld~\cite{ALFWorld}. Both are widely adopted benchmarks in the agentic RL literature~\cite{roll,verl-agent,rollart}. We train them with GRPO~\cite{shao2024deepseekmath} and DAPO~\cite{DAPO}. For all tasks, we set the group size to 16 and use batch sizes of 256 and 1024 for Qwen3-8B and Qwen3-32B, respectively. For DAPO, the rollout stage continues until it collects the target number of trajectory groups with non-zero reward variance. We adjust the maximum number of concurrent trajectories at each step based on the previous step to speed up the collection of valid trajectories. We train the 8B and 32B models on 16 and 48 dedicated GPUs, with (Rollout, Training) allocations of (8, 8) and (16, 32). Rollout TP is 1 (8B) and 4 (32B), while training parallelism (TP, PP, CP) is (4, 1, 1) and (8, 1, 2). We cap borrowed serving GPUs at 16 (8B) and 64 (32B), and set the per-device rollout batch size to 16 to avoid contention (see Appendix~\appsecref{app:rollout_profiling_thpt}). 

\noindent\textbf{Cluster Setup.} We run agentic RL training on an H800 cluster with up to 48 GPUs and online serving on a separate H800 cluster with 64 GPUs. Within each cluster, nodes are connected via 400~Gbps InfiniBand. Due to operational constraints, our evaluation uses a 200~Gbps Ethernet link for cross-cluster communication. We further evaluate communication efficiency under different link bandwidths in \S\ref{subsec:eval-transfer-engine}. We use \texttt{NCCL} and \texttt{Mooncake Store}~\cite{qin2024mooncake} for intra- and cross-cluster weight transfer. 
For the serving cluster, we pre-deploy Qwen3-8B and Qwen3-32B as rollout models, which are used by 40\% and 19\% of agentic RL training jobs in our cluster, respectively, and together cover most workloads. We co-locate them with similarly sized heterogeneous serving models (Qwen3-8B with Qwen2.5-7B, Qwen3-32B with Qwen2.5-32B) because they use compatible parallelism configurations.
We set the prefill-to-decoding instance ratio to 1:3. We replay the 24-hour online serving trace from Microsoft~\cite{stojkovic2025dynamollm}, and run a load-balancing policy to route online serving requests.

\begin{figure*}[t]
  \centering
  \begin{subfigure}[b]{0.24\linewidth}
    \centering
    \includegraphics[width=\linewidth]{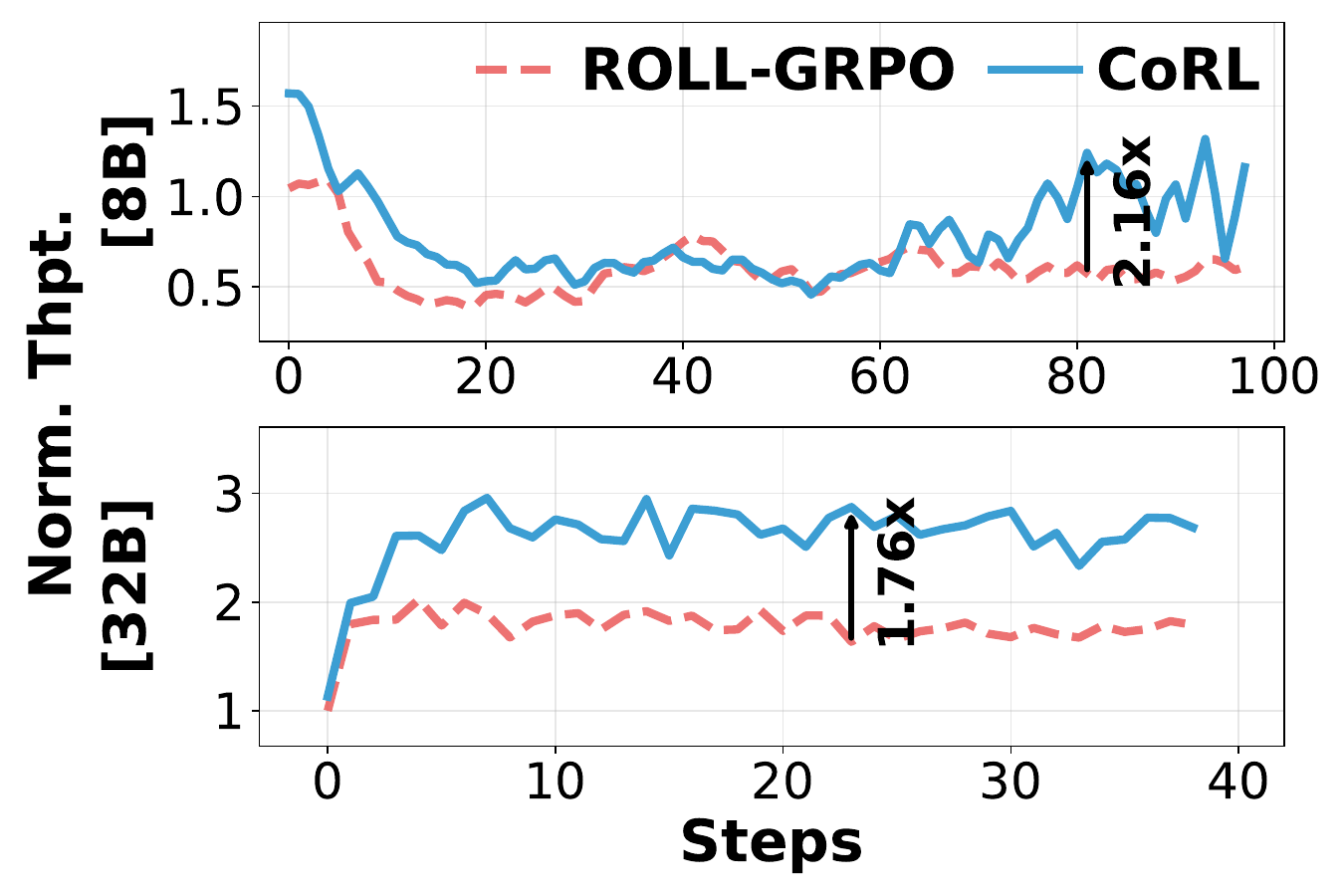}
    \caption{One-Off w/ GRPO.}
    \label{subfig:grpo_thpt}
  \end{subfigure}\hfill
  \begin{subfigure}[b]{0.24\linewidth}
    \centering
    \includegraphics[width=\linewidth]{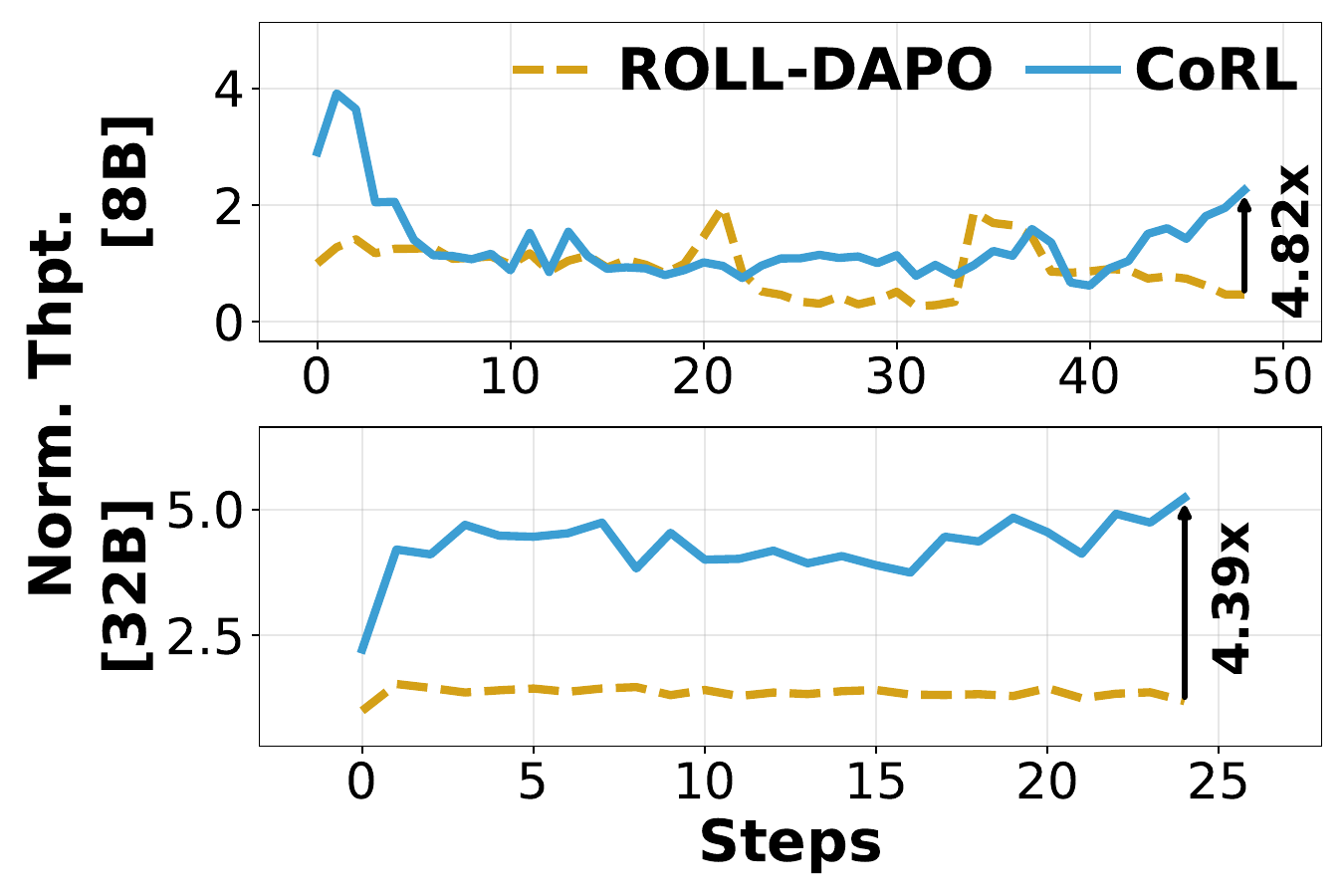}
    \caption{One-Off w/ DAPO.}
    \label{subfig:dapo_thpt}
  \end{subfigure}\hfill
  \begin{subfigure}[b]{0.24\linewidth}
    \centering
    \includegraphics[width=\linewidth]{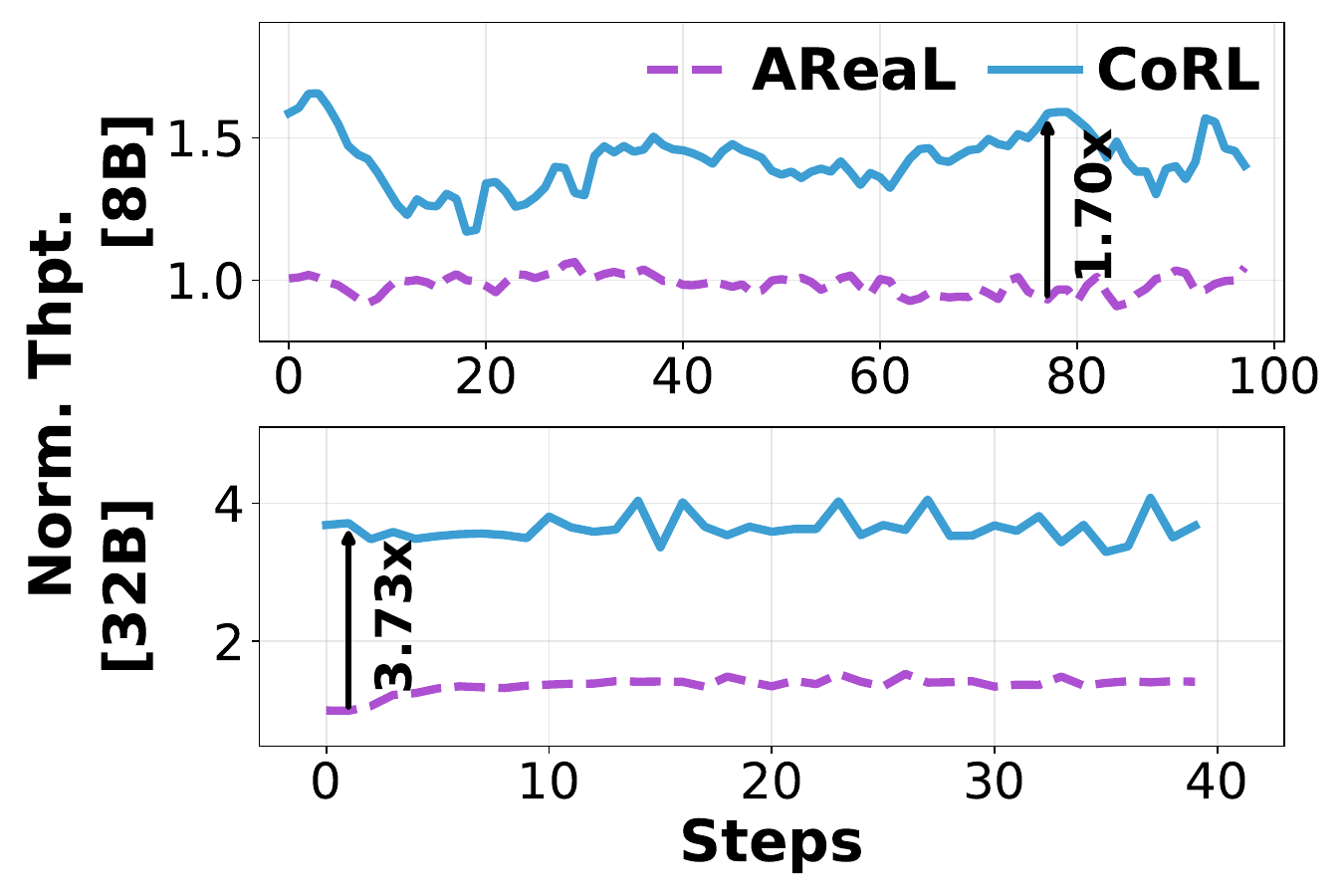}
    \caption{Fully-Async w/ AReaL.}
    \label{subfig:areal_thpt}
  \end{subfigure}\hfill
  \begin{subfigure}[b]{0.24\linewidth}
    \centering
    \includegraphics[width=\linewidth]{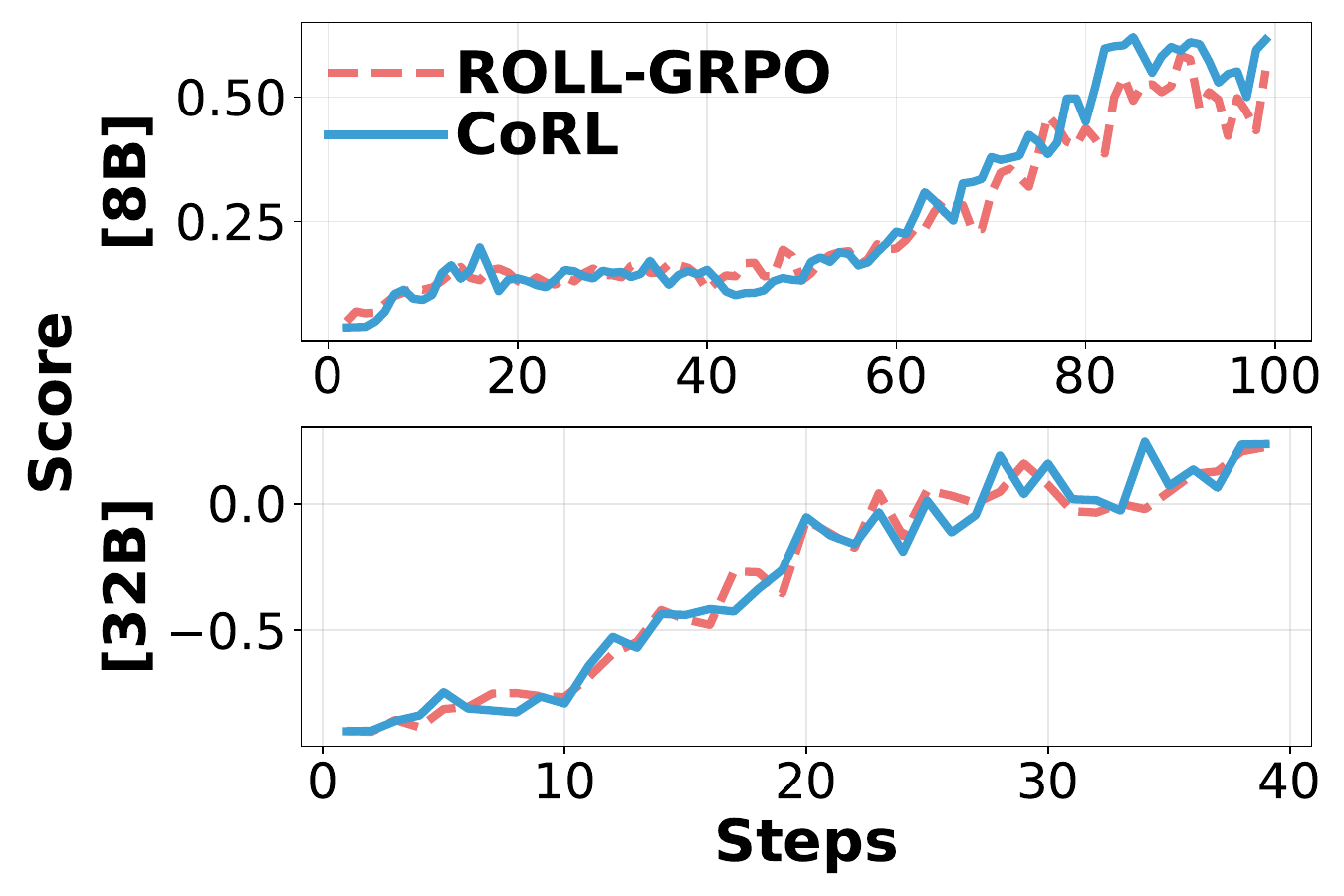}
    \caption{GRPO Critic Scores.}
    \label{subfig:grpo_score}
  \end{subfigure}\hfill
  \vspace{-10pt}
  \caption{
    (a)-(c) \SystemName{}'s end-to-end throughput improvements compared with baselines, for each baseline we run 8B and 32B model. The data are normalized to the baseline's first step.
    (d) End-to-end critic scores for 8B and 32B models using GRPO.
}
  \vspace{-5pt}
  \label{fig:e2e-thpt}
\end{figure*}

\begin{figure}[!ht]
  \centering
  \begin{subfigure}[b]{0.48\linewidth}
    \centering
    \includegraphics[width=\linewidth]{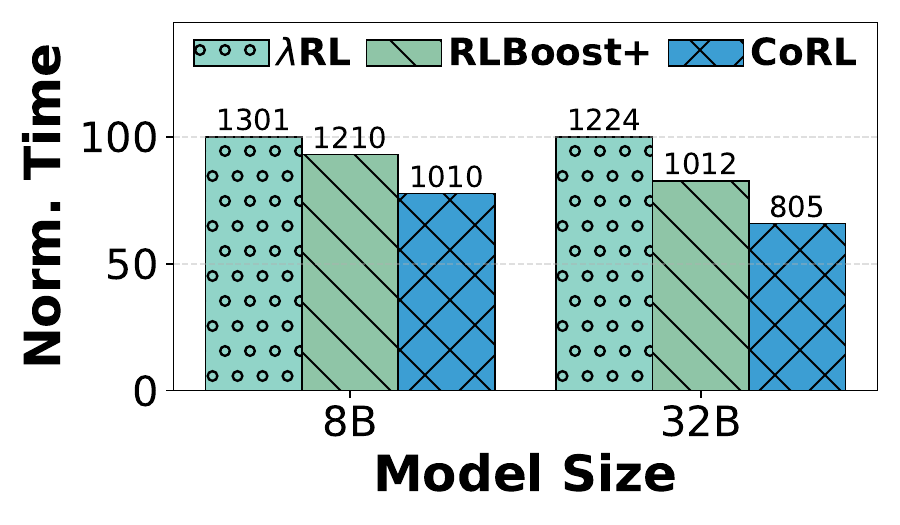}
    \caption{Elastic Baselines.}
    \label{fig:baseline_time_compare}
  \end{subfigure}\hfill
  \begin{subfigure}[b]{0.48\linewidth}
    \centering
    \includegraphics[width=\linewidth]{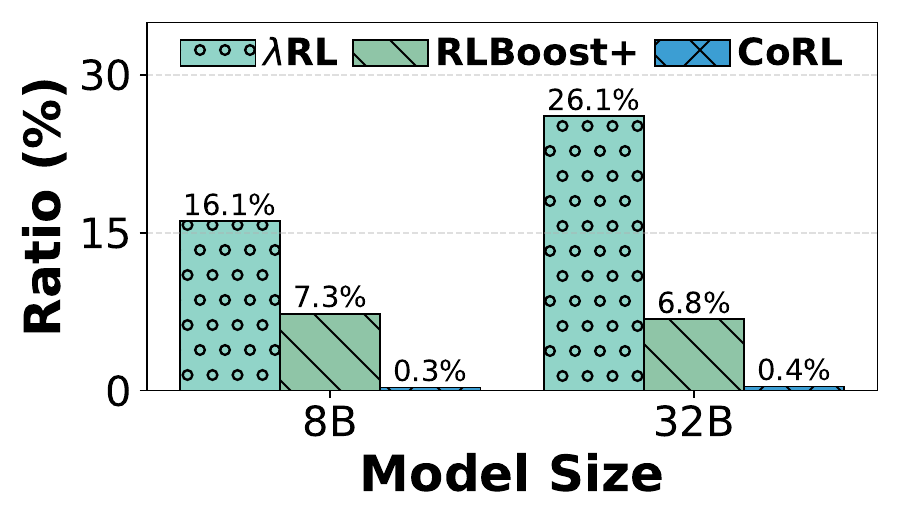}
    \caption{Allocation Overhead.}
    \label{fig:baseline_allocation_compare}
  \end{subfigure}
  \vspace{-10pt}
  \caption{End-to-end evaluation. (a) Rollout time and (b) Allocation overhead compared with elastic baselines.}
  \vspace{-10pt}
  \label{fig:micro_benchmarks_e2e}
\end{figure}

\noindent\textbf{Baselines and Training Recipe.} 
We evaluate \SystemName{} against state-of-the-art RL post-training systems.
\SystemName{} is agnostic to the choice of RL algorithms. 
Due to the substantial overhead of agentic RL, we adopt one-step off-policy training~\cite{deepcoder2025} for most baselines and additionally evaluate fully asynchronous training with AReaL~\cite{areal}.
We first compare \SystemName{} with three resource-fixed baselines in an end-to-end training evaluation:
(1)~\textbf{ROLL-GRPO}, which trains with GRPO~\cite{shao2024deepseekmath} on ROLL~\cite{roll}~\footnote{We choose ROLL over veRL~\cite{verl} for better support of agentic tasks.};
(2)~\textbf{ROLL-DAPO}, which trains with DAPO~\cite{DAPO} on ROLL; and
(3)~\textbf{AReaL}, a fully asynchronous RL system.
On \SystemName{} and all these baselines, we train until model convergence.
Specifically, for GRPO and AReaL, we train Qwen3-8B and Qwen3-32B for 100 and 40 steps, and for DAPO we use 50 and 25 steps, respectively.
We also compare rollout performance with two elastic baselines that leverage additional compute:
(4)~\textbf{$\lambda$RL}, a serverless RL baseline inspired by existing commercial products~\cite{openpipe_serverless_rl,thinkingmachines_tinker}, where the number of serverless GPUs matches the maximum available spot GPUs and the function timeout is set to 15 minutes~\cite{alibabacloud_fc_gpu_function}; and
(5)~\textbf{RLBoost+}, a strengthened RLBoost~\cite{rlboost} that uses the same training and rollout GPU allocation as ROLL, but additionally utilizes spot instances in one-step off-policy mode.
To capture the impact of spot GPU availability, we reproduce a 2-hour window characterized by high resource volatility from the traces (see Appendix~\appsecref{app:spot-instance-trace}) and run elastic baselines within this window.

\noindent\textbf{Metrics.} We measure training throughput as the total number of input and output tokens processed per global step divided by the step time~\cite{RhymeRL, StreamRL}. We use the average critic score as the accuracy metric. For serving, we set P99 SLOs for (TPOT, TTFT) to (150\,ms, 500\,ms) for Qwen2.5-7B and (450\,ms, 1000\,ms) for Qwen2.5-32B~\cite{distserve, BLITZSCALE}.

\subsection{End-to-End Evaluation}
\label{subsec:end-to-end-evaluation}
\noindent\textbf{Model Convergence.} We report the critic scores over the training in \autoref{subfig:grpo_score}. It shows that \SystemName{} preserves the model convergence due to its algorithm-agnostic design, achieving final scores nearly identical to the baselines.

\noindent\textbf{End-to-End Throughput.} 
\autoref{fig:e2e-thpt}a-c compares the end-to-end throughput against the baselines with different model and tasks.
Compared with ROLL, \SystemName{} achieves average throughput improvements of 1.31$\times$ and 1.46$\times$ with GRPO~(see \autoref{subfig:grpo_thpt}), the maximum throughput increase is 2.16$\times$ and 1.76$\times$.
The advantage of \SystemName{} becomes more pronounced under the DAPO algorithm, where average throughput increases by 1.42$\times$ and 3.31$\times$ with a maximum of 4.82$\times$ and 4.39$\times$~(see \autoref{subfig:dapo_thpt}).
Compared with AReaL, \SystemName{} achieves 1.44$\times$ and 2.69$\times$ higher throughput on average~(see \autoref{subfig:areal_thpt}).
Although AReaL eliminates GPU idle time by continuously generating trajectories without waiting for training to complete, by expanding effective GPU capacity through cooperative elasticity, \SystemName{} provides gains orthogonal to asynchronous execution.

The throughput gains are driven by three main design elements.
(1) \emph{Cooperative elasticity via co-serving executor}: by harvesting idle serving GPUs, \SystemName{} effectively expands the rollout GPU pool. With 16 additional serving GPUs, rollout time decreases by $1.69\times$ (see Appendix~\appsecref{app:scalability}), confirming that the co-serving executor successfully converts serving slack into rollout capacity. (2) \emph{Elastic rollout scheduler}: DAPO’s redundant sampling can launch up to $5.7\times$ the base batch size (\autoref{subfig:elastic}). The resource-fixed baseline cannot absorb this burst and suffers severe contention, whereas \SystemName{}’s scheduler dynamically offloads excess trajectories to serving GPUs. The scheduler’s turn-wise routing and KVC affinity together improve rollout efficiency by up to $1.48\times$ (\autoref{tab:ablation_rollout_scheduler}). (3) \emph{Weight transfer engine}: cross-cluster synchronization completes within 21\,s for Qwen3-32B (\S\ref{subsec:eval-transfer-engine}), preventing weight updates from becoming a bottleneck between steps. Moreover, \SystemName{} eliminates the allocation overhead that plagues elastic baselines (<0.5\% vs.\ 6.8--26.1\% for RLBoost+ and $\lambda$RL, \autoref{fig:baseline_allocation_compare}). Across all experiments, \SystemName{} meets the serving P99 latency SLOs (\S\ref{subsec:analysis-co-serving-executor}).

\noindent\textbf{Comparison with Elastic Baselines.} \autoref{fig:baseline_time_compare} compares rollout efficiency against $\lambda$RL and RLBoost+ throughout the training using GRPO. $\lambda$RL uses 16 and 32 serverless GPUs for rollouts of the 8B and 32B models, respectively, which is consistent with the maximum number of GPUs available to RLBoost+. We observe that allocating more GPUs to rollouts yields up to \(1.31\times\) and \(1.23\times\) speedups over ROLL for Qwen3-8B and Qwen3-32B, respectively. Although RLBoost+ experiences fluctuating spot GPU availability, its resource allocation changes less frequently than that of $\lambda$RL, which reallocates resources at a fixed 15-minute lease interval. As a result, RLBoost+ achieves \(1.41\times\) and \(1.48\times\) speedups over ROLL for Qwen3-8B and Qwen3-32B, respectively, although its gains are still constrained by the instability of spot GPUs. \SystemName{} further reduces rollout time by \(1.20\times\) and \(1.26\times\) compared to RLBoost+. This improvement suggests that opportunistically leveraging serving GPUs provides more abundant and stable compute for rollouts while avoiding frequent preemption, thereby shortening rollout time.

\noindent\textbf{Allocation Overhead.} We further analyze the allocation overhead of each elastic scheme. We define \emph{preempted GPU time} as the number of preempted GPUs multiplied by the per-preemption recovery time, normalized by total GPU time. As shown in \autoref{fig:baseline_allocation_compare}, $\lambda$RL incurs the highest overhead (up to 26.1\%), because serverless GPUs are leased for fixed durations often shorter than a complete rollout, leading to frequent preemptions and restarts. RLBoost+ experiences less frequent preemptions on spot instances but still incurs non-negligible overhead (6.8--7.3\%). In contrast, \SystemName{} requires only a one-time model activation of at most five seconds, keeping the allocation overhead ratio below 1\%.

\begin{table}[tb]
\centering
\caption{\SystemName{} vs.\ alternative serving engines: rollout time (s) and P99 serving SLO (ms).}
\vspace{-10pt}
\resizebox{0.85\linewidth}{!}{
\begin{tabularx}{\linewidth}{l l *{3}{>{\centering\arraybackslash}X}}
\toprule
\textbf{Model} & \textbf{Method} & \textbf{Rollout} & \textbf{TTFT} & \textbf{TPOT} \\
\midrule
\multirow{4}{*}{\textbf{8B}}
& \SystemName{}           & 496.3  & 338.1   & 136.1  \\
& ServerlessLLM~\cite{fu2024serverlessllm}& --     & 314.8   & 117.8 \\
& ServerlessLLM+Rollout & 651.7  & 1166.1  & 135.6 \\
& Prism~\cite{Prism}                  & 731.7 & 973.2   & 115.4 \\
\midrule
\multirow{4}{*}{\textbf{32B}}
& \SystemName{}           & 960.1  & 837.5   & 398.1  \\
& ServerlessLLM~\cite{fu2024serverlessllm}          & --     & 716.1   & 312.8  \\
& ServerlessLLM+Rollout & 1161.8 & 2426.2  & 565.3  \\
& Prism~\cite{Prism}                  & 1301.2 & 1625.4   & 351.7  \\
\bottomrule
\end{tabularx}}
\label{tab:alt_serving_engines}
\vspace{-15pt}
\end{table}

\noindent\textbf{Comparison with Alternative Serving Engines.}
Due to the high cost of full training runs, we compare alternative serving engines at a single checkpoint (step 50 for 8B, step 20 for 32B) that exhibits moderately stable rollout time and fluctuating serving load. Each experiment runs ten RL steps.

\noindent\underline{\textit{Bidirectional Autoscaling.}} We replace \SystemName{}'s co-serving executor with ServerlessLLM~\cite{fu2024serverlessllm}. As shown in \autoref{tab:alt_serving_engines}, ServerlessLLM achieves low serving latency without rollouts (TTFT-P99: 314.8\,ms for 8B), but severely violates SLOs when rollouts are enabled (1166\,ms and 2426\,ms). Repeated rollout model eviction and reloading during serving bursts inflates rollout time by 1.31$\times$ (8B) and 1.21$\times$ (32B), confirming that bidirectional autoscaling cannot safely harvest serving slack under fluctuating traffic.

\noindent\underline{\textit{GPU Multiplexing.}} We evaluate Prism~\cite{Prism}, a heterogeneous LLM colocation engine, configured to co-schedule rollout and serving requests (rollout SLO = $4\times$ serving SLO). Prism does not provide dual-SLO admission support or prefix caching for rollouts. As shown in \autoref{tab:alt_serving_engines}, the lack of prefix caching inflates rollout time by $1.47\times$ (8B) and $1.36\times$ (32B), while the absence of dual-SLO admission support leads to TTFT SLO violations (973\,ms and 1625\,ms vs.\ 500\,ms and 1000\,ms targets). This demonstrates that generic multiplexing without RL-aware co-serving is insufficient.

\begin{table}[tb]
  \centering
  \caption{\textbf{[Co-Serve Executor]}. The impact of memory sharing policy for rollout and SLO efficiency.}
  \label{tab:latency}
  \vspace{-10pt}
  \small
  \begin{adjustbox}{width=0.95\linewidth}
    \begin{tabular}{llccccc}
      \toprule
      \multirow{2}{*}{\textbf{Model}} &
      \multirow{2}{*}{\textbf{Policy}} &
      \multicolumn{1}{c}{\textbf{Rollout}} &
      \multicolumn{2}{c}{\textbf{TTFT (ms)}} &
      \multicolumn{2}{c}{\textbf{TPOT (ms)}} \\
      \cmidrule(lr){3-3}\cmidrule(lr){4-5}\cmidrule(lr){6-7}
      & & Time (s) & P95 & P99 & P95 & P99 \\
      \midrule
      \multirow{3}{*}{Qwen3-8B}
        & Static Partition     & 776.7  & 299.5 & 330.95 & 1462.5 & 1488.2 \\
        & + Memory Preemption & 591.7  & 348.8 & 517.9  & 153.6  & 162.8  \\
        & + Prefix Caching    & 469.3  & 318.5 & 333.6  & 173.2  & 186.3  \\
      \midrule
      \multirow{3}{*}{Qwen3-32B}
        & Static Partition     & 1310.5 & 595.6 & 603    & 6835.9 & 7021.1 \\
        & + Memory Preemption & 959.7  & 596.7 & 651.3  & 493.5  & 503.8  \\
        & + Prefix Caching    & 920.4  & 607.2 & 640.6 & 477.9  & 483.2  \\
      \bottomrule
    \end{tabular}
  \end{adjustbox}
  \vspace{-10pt}
\end{table}

\subsection{Analysis of Co-serving Executor}
\label{subsec:analysis-co-serving-executor}
To analyze the efficiency of the co-serving executor, we use the end-to-end setups for Qwen3-8B and Qwen3-32B and run GRPO for five steps. We utilize the same model checkpoint and serving traffic in \autoref{tab:alt_serving_engines}. 

\noindent\textbf{Preemptive Dynamic Memory Sharing Policy.} This policy incorporates two memory-sharing optimizations, namely \emph{memory preemption} and \emph{prefix caching}. We run rollouts on serving GPUs via spatial GPU sharing and measure end-to-end rollout time, along with serving P95 and P99 TTFT and TPOT, to quantify rollout efficiency and SLO compliance. We first evaluate memory preemption. As a \emph{static partition} baseline, we split GPU memory evenly between serving and rollout LLMs. Compared to \emph{static partitioning}, memory preemption reduces P99 TPOT latency by \(9.14\times\) for Qwen3-8B and \(13.94\times\) for Qwen3-32B, indicating that it absorbs bursty serving-side memory demand and reduces tail latency.

Second, we enable \emph{prefix caching} atop memory preemption. This reduces rollout time by \(1.26\times\) for Qwen3-8B and \(1.04\times\) for Qwen3-32B, with only a slight increase in serving tail latency. These results indicate that prefix caching mainly improves rollout throughput, but does not by itself ensure serving SLO compliance. Overall, the memory sharing policy effectively limits tail-latency inflation, but without explicit SLO-aware scheduling, P99 latency still misses our target.

\begin{figure}[tb]
  \centering
    \centering
    \includegraphics[width=0.95\linewidth]{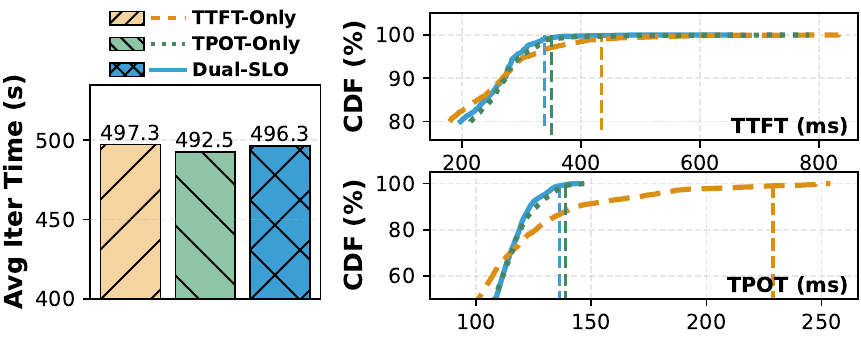}
  \caption{Analysis of Dual-SLO Admission Controller.}
  \label{fig:dual_slo}
\end{figure}

\begin{figure}[tb]
  \centering
    \begin{subfigure}[b]{0.48\linewidth}
    \centering
    \includegraphics[width=\linewidth]{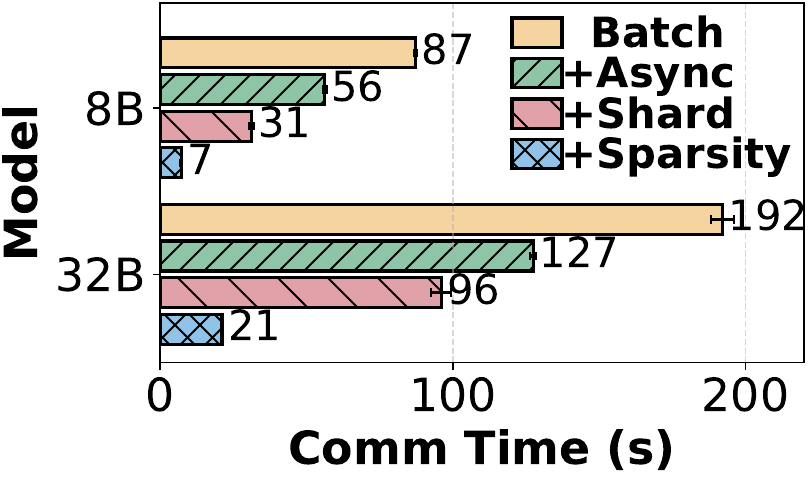}
    \caption{Weight transfer overhead.}
    \label{fig:comm-weight-transfer}
  \end{subfigure}\hfill
  \begin{subfigure}[b]{0.48\linewidth}
    \centering
    \includegraphics[width=\linewidth]{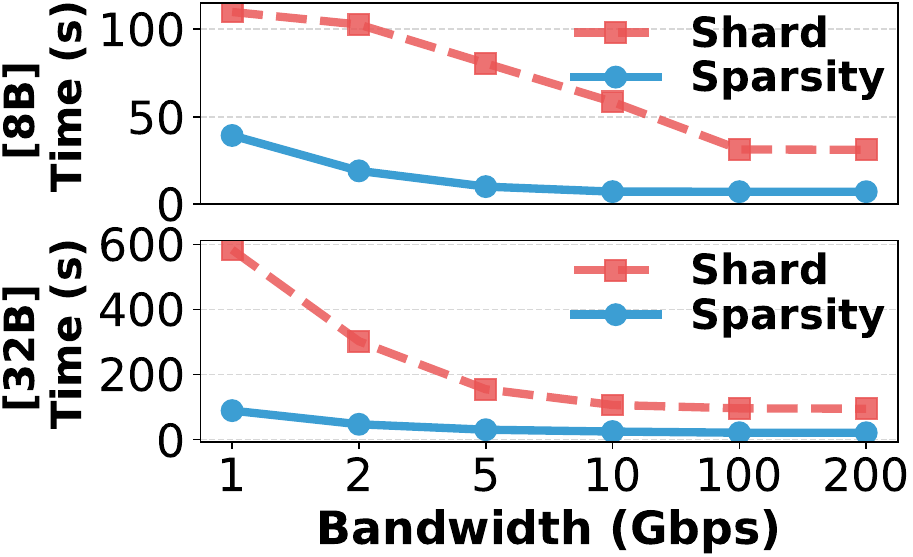}
    \caption{Sensitivity to bandwidth.}
    \label{fig:varying-bandwidth}
  \end{subfigure}\hfill
  \vspace{-8pt}
\caption{\textbf{[Transfer Engine]}. (a) Cross-cluster weight transfer time under different optimizations; each optimization is additive over the previous one. (b) Sensitivity of shard-aware and sparsity-aware transfer of different LLMs to cross-cluster bandwidth.}
\label{fig:comm-opt}
\vspace{-10pt}
\end{figure}

\noindent\textbf{Dual-SLO Admission Controller.} This controller compares the current serving TTFT and TPOT against their SLO targets and time-multiplexes between serving traffic and rollout requests to maintain SLO compliance. We evaluate three policies under the same setup in \autoref{fig:dual_slo}: \emph{TTFT-only}, \emph{TPOT-only}, and \emph{Dual-SLO}. \emph{TTFT-only} enforces only the TTFT SLO, while \emph{TPOT-only} enforces only the TPOT SLO. The average step time is similar across all three policies, suggesting that the policy choice has limited impact on rollout time. Compared to the single-objective baselines, \emph{Dual-SLO} consistently achieves lower P99 tail latency on both metrics. It reduces P99 TPOT by 3.4\% and 76.3\% relative to \emph{TPOT-only} and \emph{TTFT-only}, respectively, and reduces P99 TTFT by 3.7\% and 28.4\% over the same baselines. Overall, \emph{Dual-SLO} explicitly provides guarantees for both TTFT and TPOT while preserving rollout efficiency. We further explore the sensitivity of the co-serving executor to prefix caching lease time (Appendix~\appsecref{app:sensivity-to-lease}) and online serving load (Appendix~\appsecref{app:sensivity-to-serving}), demonstrating its robustness.

\subsection{Effectiveness of Transfer Engine}
\label{subsec:eval-transfer-engine}
We evaluate the transfer engine using 16 GPUs for training and 16 GPUs for serving, and measure cross-cluster communication overhead over three RL steps. Unless otherwise specified, the training and serving clusters are connected by a 200~Gbps Ethernet link.

\noindent\textbf{Analysis of Weight Transfer Optimizations.} \autoref{fig:comm-weight-transfer} quantifies the cross-cluster communication efficiency of \emph{asynchrony}, \emph{shard-awareness}, and \emph{sparsity-awareness} for Qwen3-8B and Qwen3-32B. As a baseline (\emph{batch}), training workers all-gather model weights and transmit the entire model, and then serving workers pull model weights from relay workers into GPU memory. \emph{Asynchronous} transfer streams parameters at bucket granularity and pipelines publishing with pulling, reducing end-to-end communication time by {1.6$\times$} (Qwen3-8B) and {1.5$\times$} (Qwen3-32B). \emph{Shard-awareness} enables each training worker to publish only its local shard, and each serving worker to pull only the shards it hosts, further reducing communication time by 1.8$\times$ (Qwen3-8B) and 1.3$\times$ (Qwen3-32B). \emph{Sparsity-awareness} provides the largest gain, reducing communication time by {4.4$\times$} (Qwen3-8B) and {4.6$\times$} (Qwen3-32B). Overall, our optimized cross-cluster weight transfer engine reduces end-to-end communication time by {12.4$\times$} for Qwen3-8B and {9.1$\times$} for Qwen3-32B, bringing cross-cluster transfer down to tens of seconds. Detailed timeline breakdown is provided in Appendix~\appsecref{app:timeline-breakdown}.

\begin{table}[tb]
\centering
\caption{\textbf{[Rollout Scheduler]}. The speedup of the elastic rollout scheduler on rollout time.}
\vspace{-10pt}
\resizebox{0.85\linewidth}{!}{
\begin{tabularx}{\linewidth}{l *{3}{>{\centering\arraybackslash}X}}
\toprule
\textbf{Policy} & \textbf{Qwen3-8B} & \textbf{Qwen3-32B}\\
\midrule
\textbf{~~Baseline} & 1.00$\times$ & 1.00$\times$ \\
\midrule
\textbf{+~Turn-Wise Routing} & 1.11$\times$ & 1.08$\times$ \\
\midrule
\textbf{+~KVC Affinity} & 1.16$\times$ & 1.48$\times$ \\
\bottomrule
\end{tabularx}}
\label{tab:ablation_rollout_scheduler}
\vspace{-15pt}
\end{table}

\noindent\textbf{Sensitivity to Cross-Cluster Bandwidth.}
We study the sensitivity of shard-aware and sparsity-aware transfer to cross-cluster bandwidth by throttling link capacity from 200\,Gbps to 1\,Gbps using \texttt{tc} command, and measuring end-to-end communication overhead for Qwen3-8B (\autoref{fig:varying-bandwidth}, top) and Qwen3-32B (\autoref{fig:varying-bandwidth}, bottom). For Qwen3-8B, shard-aware transfer increases from 31\,s (200\,Gbps) to 109\,s (1\,Gbps), while sparsity-aware transfer stays within 7--39\,s, delivering a $2.8\times$--$4.4\times$ speedup. For Qwen3-32B, shard-aware transfer grows from 96\,s to 584\,s, whereas sparsity-aware transfer remains within 21--89\,s ($4.6\times$--$6.6\times$ faster). Overall, sparsity-awareness flattens the scaling curve by reducing transferred bytes. Even at 5\,Gbps, transfer completes within 10\,s (Qwen3-8B) and 30\,s (Qwen3-32B). Because cross-cluster transfer is overlapped with training and intra-cluster communication, it increases training time by at most 5\%. 

\begin{figure}[t]
  \centering
  \begin{subfigure}[b]{0.48\linewidth}
    \centering
    \includegraphics[width=\linewidth]{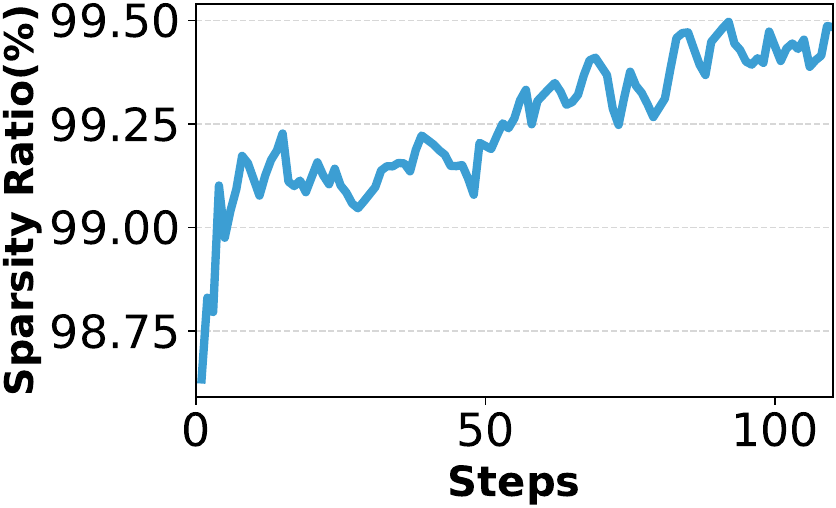}
    \caption{Sparsity Ratio [8B].}
    \label{fig:sparsity-over-iter}
  \end{subfigure}\hfill
  \begin{subfigure}[b]{0.48\linewidth}
    \centering
    \includegraphics[width=\linewidth]{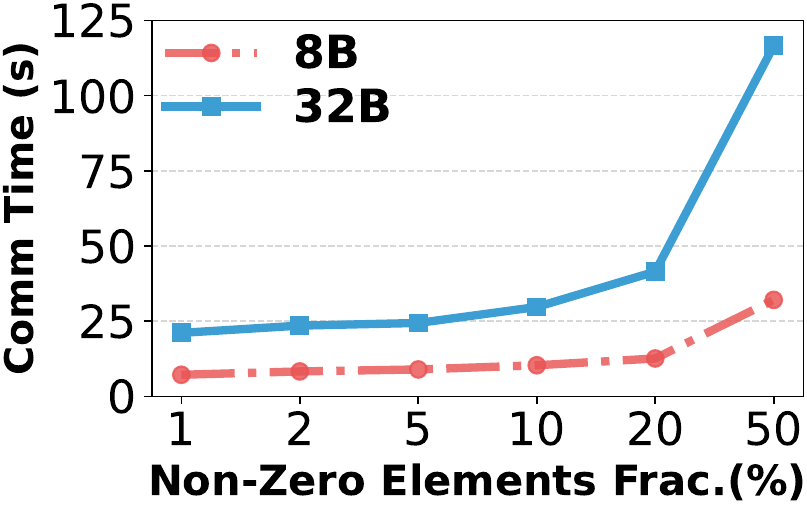}
    \caption{Sensitivity to Sparsity.}
    \label{fig:sensivity-to-sparsity}
  \end{subfigure}
  \caption{\textbf{[Analysis of Sparsity]}. (a) The sparsity of weight differentials across steps for Qwen3-8B. (b) The sensitivity of transfer engine to sparsity.}
  \label{fig:two_side_by_side}
\end{figure}

\noindent\textbf{Analysis of Weight Differential Sparsity.}
\autoref{fig:sparsity-over-iter} shows that the weight-differential sparsity for Qwen3-8B stays around 99\% throughout training. This indicates that sparsity is persistent rather than a transient artifact of a particular training phase.   We next vary the non-zero fraction in \autoref{fig:sensivity-to-sparsity} for Qwen3-8B and Qwen3-32B to evaluate the sensitivity of our sparsity-aware transfer engine to the sparsity ratio. As the non-zero fraction increases, communication overhead rises and the benefit of sparse transfer diminishes. Beyond \(\sim\)20\%, sparse-format metadata (e.g., indices) and (de)sparsification overhead begin to offset the reduction in transmitted weights. In our workloads, the measured non-zero fraction remains well below this threshold, enabling consistently efficient weight propagation.

\subsection{Effectiveness of Rollout Scheduler.} 
We follow the end-to-end setups and evaluate the elastic rollout scheduler using Qwen3-8B and Qwen3-32B with GRPO algorithm for the first five RL steps. \autoref{tab:ablation_rollout_scheduler} analyzes the contribution of two heuristics adopted by the rollout scheduler. As a baseline, we pin each trajectory to a fixed rollout worker for its entire lifetime. This static assignment leads to load imbalance because trajectory runtimes vary widely. As a result, enabling turn-wise routing reduces rollout time by \(1.11\times\) (Qwen3-8B) and \(1.08\times\) (Qwen3-32B), demonstrating the benefit of fine-grained, flexible routing. Adding KVC affinity yields further improvements, bringing the cumulative speedup to \(1.16\times\) for Qwen3-8B and \(1.48\times\) for Qwen3-32B. This is because larger models incur heavier prefill costs, making KV reuse more effective at reducing per-turn overhead. Overall, these simple heuristics substantially improve rollout efficiency. The scheduler adds negligible overhead (at most 10~ms), suggesting that it scales well.

\section{Related Works}
\noindent\textbf{Agentic RL Training Systems.} Many systems optimize conventional single-turn RL post-training~\cite{hu2024openrlhf,sheng2024hybridflow,deepspeedchat,Harper_NeMo_a_toolkit,zhong2024rlhfuseefficientrlhftraining,asyncflow,RhymeRL,rollpacker,StreamRL,rlboost,seer-sd,best-of-n-sd}. The rise of agentic LLMs has also motivated agentic RL training frameworks~\cite{verl-agent,roll,rollart,skyrl-agent,ALFWorld,RollMux}. However, these agentic systems typically assume fixed resources. A few elastic RL systems exploit spot instances~\cite{rlboost} or serverless GPUs~\cite{openpipe_serverless_rl,thinkingmachines_tinker}. \SystemName{} explores underutilized serving GPUs for resource elasticity and is complementary to these elastic approaches.

\noindent\textbf{Serving GPU Sharing.} 
GPU multiplexing has been widely studied. 
Prior DL systems~\cite{xiao2020antman,zhao2022multi,icefrog,li2023alpaserve,han2022microsecond} interleave multiple models on the same GPUs to improve utilization.
Recent LLM serving systems extend multiplexing to co-serve multiple LLMs: systems targeting homogeneous models~\cite{ConServe} share a single KVC layout across co-located workloads, while heterogeneity-aware systems~\cite{Prism,Aegaeon,duan2024muxserve,seallm,llm-multitasking} multiplex workloads with comparable SLO targets.
Lyra~\cite{Lyra} loans idle serving servers to training at machine granularity.
\SystemName{} targets a different setting: co-locating heterogeneous models with asymmetric SLO requirements on the same GPU.



\noindent\textbf{Sparsity-based Optimization.} Recent systems including Check-N-Run~\cite{eisenman2022check} and LowDiff~\cite{freqcheckpoint} leverage the sparsity to reduce storage and checkpoint overhead. Many communication optimization systems~\cite{StellaTrain,OmniReduce,ZEN,QSGD} exploit gradient sparsity to cut communication cost with lossy compression. Inspired by these works, we observe the sparsity in the weight differential of RL training and leverage it to reduce the communication overhead with lossless compression.

\noindent\textbf{Cycle Stealing.} Harvesting idle resources across workloads is a well-studied concept. Cycle stealing~\cite{CycleStealing} demonstrated that beneficiaries gain unbounded benefit from donors' idle cycles with only slight donor penalty. In the GPU era, Ekya~\cite{Ekya} balances inference and continuous retraining on edge GPUs, and Lyra~\cite{Lyra} loans idle serving GPUs to training jobs. \SystemName{} extends this philosophy to co-serving heterogeneous LLMs, additionally handling incompatible KVC layouts, prefix caching contention, and cross-cluster weight synchronization.

\section{Conclusion}
In this paper, we present \SystemName{}, a cooperative, elastic post-training system that opportunistically harvests serving GPUs to accelerate agentic RL training. \SystemName{} combines (i) a co-serving executor for SLO-safe compute and memory sharing, (ii) a cross-cluster weight transfer engine for low-overhead weight synchronization, and (iii) an elastic rollout scheduler that efficiently realizes the cooperative elasticity. Extensive experiments show that \SystemName{} improves training efficiency over baselines while preserving serving SLOs.

\bibliographystyle{ACM-Reference-Format}
\bibliography{reference}

\clearpage

\ifsubmit
\else
  \appendix
  \section*{Appendix}

\section{Rollout Concurrency Profiling}
\label{app:rollout_profiling_thpt}
\SystemName{} profiles and caps rollout concurrency on dedicated rollout GPUs to avoid excessive KV cache (KVC) memory pressure. \autoref*{fig:throughput_real_h800} shows the rollout throughput of Qwen3-8B with a 32K context length under different per-GPU batch sizes. Throughput increases with concurrency up to a batch size of 16, after which it saturates. Increasing concurrency beyond this point increases rollout latency due to memory contention and KVC fragmentation, which in turn degrades effective throughput. Therefore, unless otherwise specified, we cap the maximum number of concurrent rollout requests per dedicated rollout GPU at 16.

\label{app:workload-characterization}
\begin{figure}[htb]
  \centering
    \centering
    \includegraphics[width=0.7\linewidth]{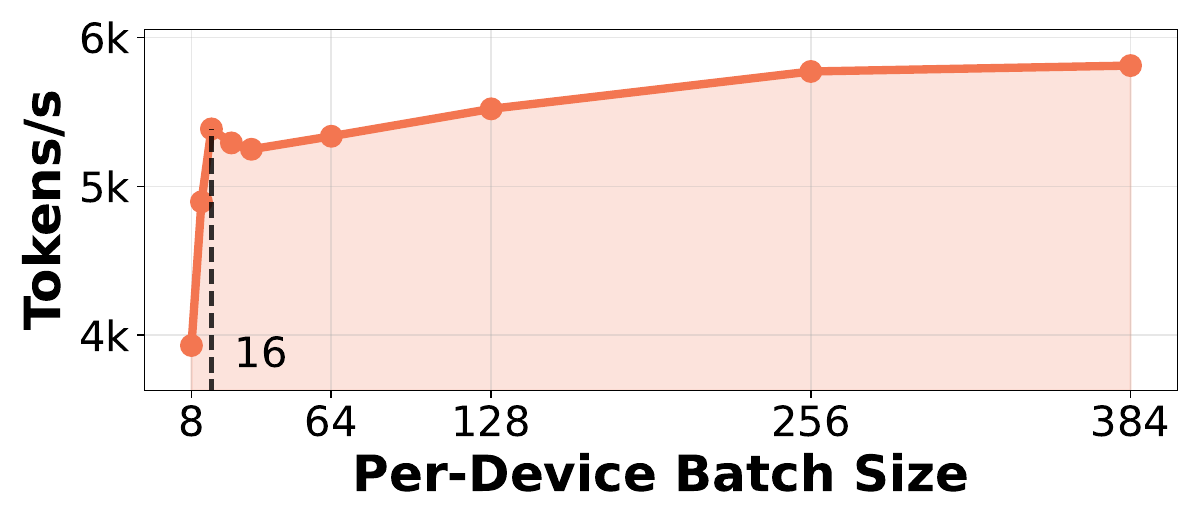}
    \vspace{-10pt}
  \caption{The system throughput with different per-device batch sizes. [Qwen3-8B/32K]}
  \label{fig:throughput_real_h800}
  \vspace{-10pt}
\end{figure}

\section{Spot instance trace}
\label{app:spot-instance-trace}

We extract the spot-instance traces for the 8B model from Seg.B in Figure 8(a) and for the 32B model from Seg.B in Figure 9 of the RLBoost paper \cite{rlboost}. \autoref*{fig:rlboost_trace} shows the variations in the number of preemptible and reserved GPUs over time. In the experiments, we used these traces to evaluate RLBoost+’s performance.

\begin{figure}[htb]
  \centering
    \centering
    \includegraphics[width=0.9\linewidth]{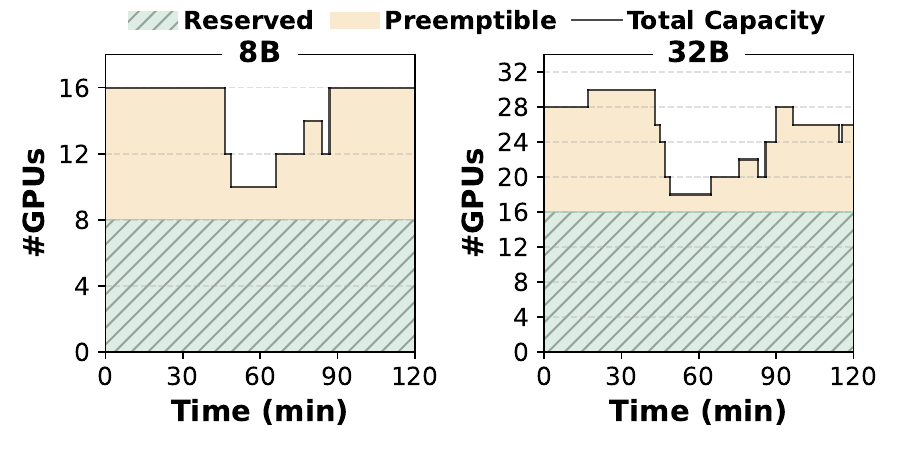}
  \caption{RLBoost trace used for 8B and 32B models.}
  \label{fig:rlboost_trace}
  \vspace{-10pt}
\end{figure}

\section{Sensitivity to caching lease}
\label{app:sensivity-to-lease}
We set the lease time for prefix cache of rollout in the memory sharing policy. \autoref*{tab:lease_impact} evaluates how the prefix cache lease time affects rollout efficiency and serving SLOs for Qwen3-8B. The rollout time is largely insensitive to the lease time, suggesting that longer cache persistence provides limited additional benefit for rollouts in this setting. In contrast, lease time has a clear impact on serving tail latency: since the Dual-SLO admission controller is designed to enforce SLO thresholds rather than explicitly minimize tail latency, a longer lease can keep more rollout KVC resident and reduce GPU memory headroom for bursty serving traffic, inflating P99 latency. Consequently, a moderate lease time offers the best trade-off, retaining most prefix cache of rollouts benefits while provide scheduling flexibility to prevent rollout prefix-cache residency from degrading serving SLOs.

\begin{table}[htb]
  \centering
  \caption{[Co-Serve Executor]. The impact of prefix-cache lease time on rollout and serving SLO.}
  \label{tab:lease_impact}
  \vspace{-8pt}
  \small
  \begin{adjustbox}{width=0.95\linewidth}
    \begin{tabular}{l c c c}
      \toprule
      \textbf{Lease (s)} & \textbf{Avg Rollout Time (s)} & \textbf{TTFT P99 (ms)} & \textbf{TPOT P99 (ms)} \\
      \midrule
      10  & 496.3 & 338.11 & 136.13 \\
      20  & 497.1 & 334.71 & 135.28 \\
      50  & 492.2 & 371.10 & 140.70 \\
      100 & 502.1 & 491.90 & 148.10 \\
      \bottomrule
    \end{tabular}
  \end{adjustbox}
  \vspace{-10pt}
\end{table}

\section{Sensitivity to Serving Traffic.}
\label{app:sensivity-to-serving}

Serving GPUs can be repurposed for rollouts because serving demand fluctuates over time, leaving transient compute and memory headroom. To understand how \SystemName{} behaves as this headroom shrinks, we scale the serving request arrival density and measure both rollout efficiency (average rollout time) and serving QoS (TTFT/TPOT P99). \autoref*{tab:density_impact} shows that higher serving density increases resource contention, which in turn prolongs rollouts and degrades serving tail latency.

Two additional observations stand out. First, TTFT is more sensitive to increasing load than TPOT, suggesting that contention primarily hurts the prefill, while per-token generation is comparatively less affected. Second, the larger model exhibits more stable rollout time as serving density increases, but its serving tail latencies still rise with load, indicating that compute/memory interference persists and must be managed by the co-serving executor. Overall, the results confirm the expected trade-off under cooperative elasticity: as serving load grows, \SystemName{} gradually reduces effective rollout capacity while keeping serving performance degradation bounded rather than causing sharp SLO violations.

\begin{table}[htb]
  \centering
  \caption{[Co-Serve Executor]. The impact of serving traffic density on rollout and serving SLO.}
  \label{tab:density_impact}
  \vspace{-8pt}
  \small
  \begin{adjustbox}{width=0.95\linewidth}
    \begin{tabular}{l c c c c}
      \toprule
      \textbf{Model} & \textbf{Density} & \textbf{Avg Rollout Time (s)} & \textbf{TTFT P99 (ms)} & \textbf{TPOT P99 (ms)} \\
      \midrule
      \multirow{3}{*}{Qwen3-8B}  & 1   & 496.3 & 338.1 & 136.1 \\
                               & 1.5 & 511.7 & 380.2  & 149.0  \\
                               & 2   & 569.9 & 459.1  & 150.1  \\
      \midrule
      \multirow{3}{*}{Qwen3-32B} & 1   & 960.1 & 837.5  & 398.1  \\
                               & 1.5 & 977.7 & 870.2  & 421.3  \\
                               & 2   & 989.9 & 899.1  & 441.2  \\
      \bottomrule
    \end{tabular}
  \end{adjustbox}
  \vspace{-10pt}
\end{table}

\section{Serving GPU Availability} 
\label{app:scalability}
The amount of underutilized capacity in the serving cluster also affects rollout performance. To quantify this effect, we measure the average rollout time of Qwen3-8B on FrozenLake using eight dedicated rollout GPUs over the first five RL steps, while varying the number of available serving GPUs. As shown in \autoref*{fig:scalability-8b}, rollout time decreases as the serving GPU quota increases. With an additional 16, 8 and 4 serving GPUs, \SystemName{} reduces rollout time by \(1.69\times\), \(1.45\times\), and \(1.26\times\), respectively. These results show that \SystemName{} can effectively harvest underutilized serving resources to reduce the rollout overhead. 
\begin{figure}[htb]
  \centering
  \includegraphics[width=0.7\linewidth]{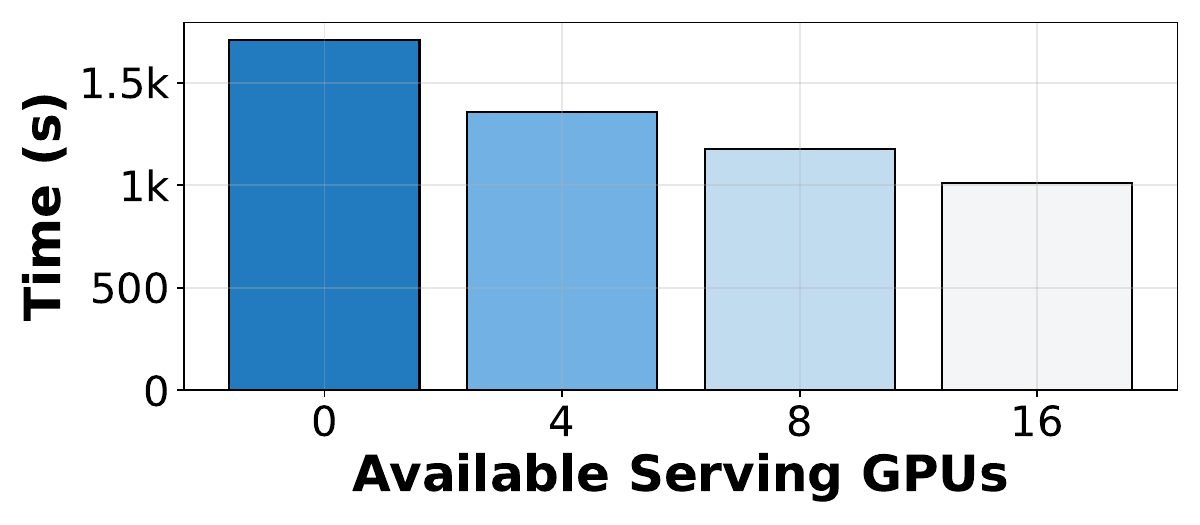}
  \caption{Sensitive Analysis of Serving GPU Availability.}
  \label{fig:scalability-8b}
  \vspace{-10pt}
\end{figure}

\section{Timeline Breakdown of Weight Transfer}
\label{app:timeline-breakdown}

\autoref*{fig:timeline-of-different-methods} provides a detailed timeline breakdown of shard-aware and sparsity-aware weight transfer for Qwen3-32B. The top timeline illustrates shard-aware transfer: on the sender side, each training worker streams $\sim$60 buckets (64\,MB each); each bucket takes 0.2--0.4\,s to push, for a total of 65 seconds. On the receiver side, serving workers pull the corresponding weight buckets from the relay and load them into GPU memory, taking 42\,s in total. Overall, shard-awareness prevents redundant weight all-gather and avoids pulling and re-sharding a full replica on each serving worker, while increasing effective parallelism by utilizing multiple cross-cluster links.

The bottom timeline shows sparsity-aware transfer for the same model. Because weight deltas are highly sparse, each bucket is much smaller: per-bucket push and pull latency drops from hundreds of milliseconds to just a few milliseconds. The remaining per-bucket overhead is dominated by (de)sparsification, specifically Dense-to-Sparse (D2S) on the training side and Sparse-to-Dense (S2D) on the serving side, which stays sub-second. As a result, end-to-end transfer time decreases to 21\,s, demonstrating the substantial benefit of exploiting weight delta sparsity for cross-cluster synchronization.

\begin{figure}[h]
  \centering
  \includegraphics[width=0.9\linewidth]{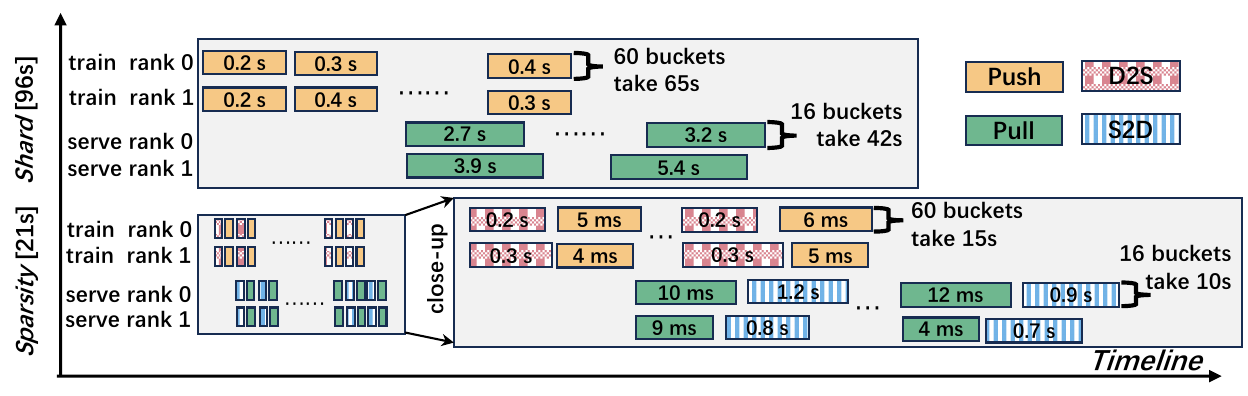}
  \caption{Timeline breakdown of shard-aware and sparsity-aware transfer for Qwen3-32B. D2S denotes the dense-to-sparse conversion, and S2D denotes the sparse-to-dense conversion.}
  \label{fig:timeline-of-different-methods}
  \vspace{-10pt}
\end{figure}
\fi

\end{document}